\documentclass{elsart}
\usepackage{amssymb}
\usepackage{newlfont}
\usepackage{graphicx,algorithmic,algorithm}
\usepackage[numbers,square,comma]{natbib}

\makeatletter
\@addtoreset{equation}{section}
\makeatother

\newcommand{\figscale}{0.7}
\newcommand{\Npoly}{N_{\mathrm{poly}}}
\newcommand{\hatD}{\hat{D}_{oo}}

\begin{document}
\begin{frontmatter}

\begin{flushright}
KEK-CP-130\\
UTCCP-P-124\\
UTHEP-458
\end{flushright}

\title{An Exact Algorithm for 
       Any-flavor Lattice QCD with Kogut-Susskind Fermion}

\author[Tsukuba]{S.~Aoki},
\author[ICRR]{M.~Fukugita},
\author[KEK]{S.~Hashimoto},
\author[Tsukuba,RCCP]{K-I.~Ishikawa},
\author[Tsukuba,RCCP]{N.~Ishizuka},
\author[Tsukuba]{Y.~Iwasaki},
\author[Tsukuba]{K.~Kanaya},
\author[KEK]{T.~Kaneko},
\author[KEK]{Y.~Kuramashi},
\author[Hiroshima]{M.~Okawa},
\author[KEK]{N.~Tsutsui},
\author[Tsukuba,RCCP]{A.~Ukawa},
\author[KEK]{N.~Yamada}, 
and \author[Tsukuba,RCCP]{T.~Yoshi\'{e}}
\collab{(JLQCD collaboration)}

\address[Tsukuba]%
{Institute of Physics, University of Tsukuba, Tsukuba, Ibaraki 305-8571, Japan}
\address[ICRR]%
{Institute for Cosmic Ray Research, University of Tokyo, Kashiwa, Chiba 277-8582, Japan}
\address[KEK]%
{High Energy Accelerator Research Organization(KEK), Tsukuba, Ibaraki 305-0801, Japan}
\address[RCCP]%
{Center for Computational Physics, University of Tsukuba, Tsukuba, Ibaraki 305-8577, Japan}
\address[Hiroshima]%
{Department of Physics, Hiroshima University, Higashi-Hiroshima, Hiroshima 739-8526, Japan}
\begin{abstract}
We propose an exact simulation algorithm for lattice QCD with dynamical 
Kogut-Susskind fermion in which the $N_f$-flavor fermion operator is 
defined as the $N_f/4$-th root of the Kogut-Susskind (KS) fermion operator. 
The algorithm is an extension of the Polynomial Hybrid Monte 
Carlo (PHMC) algorithm to KS fermions. 
The fractional power of the KS fermion operator is approximated with 
a Hermitian Chebyshev polynomial,
with which we can construct an algorithm for any number of flavors.
The error which arises from the approximation is corrected by the Kennedy-Kuti 
noisy Metropolis test. 
Numerical simulations are performed for the two-flavor case 
for several lattice parameters  in order to 
confirm the validity and the practical feasibility of the algorithm. 
In particular tests on a $16^4$ lattice with a quark mass 
corresponding to $m_{\mathrm{PS}}/m_{\mathrm{V}}\sim 0.68$ are successfully 
accomplished.
We conclude that our algorithm provides an attractive exact method for dynamical
QCD simulations with KS fermions.
\end{abstract}

\end{frontmatter}

\section{Introduction}
\label{sec:sec1}

In numerical studies of lattice QCD, advancing simulations 
including dynamical quarks is the most pressing issue to confirm 
the validity of QCD and to extract low energy hadronic properties
from it.  While including dynamical quark effects is still a difficult task,
recent developments of computational power and algorithms have enabled 
dynamical simulations of reasonable scale.  
Much efforts are thus being spent to accurately compute physical 
quantities in full QCD simulations~\cite{SESAM,SESAM_TCHIL,Eiker,CPPACS,UKQCD,JLQCD,MILC}.

A large number of these simulations are being made with Wilson-type 
fermion action using the Hybrid Monte Carlo (HMC) algorithm~\cite{HMC,DetailedHMC}, 
which is one of the exact dynamical fermion algorithms.  
A limitation of the HMC algorithm is that the number of flavors has to be 
even to express the fermion determinant in terms of pseudo-fermion fields.
Recently, however, 
the polynomial Hybrid Monte Carlo (PHMC) algorithm~\cite{PHMC_ORG,PHMC_Frezzotti_Jansen}
has been proposed to simulate the odd number of flavors with Wilson-type 
fermion as an exact algorithm~\cite{PHMC_Takaishi_Forcrand,PHMC_JLQCD}. 
The combination of the HMC and PHMC algorithms enables us to simulate 
the realistic world of 2+1-flavors of 
quarks with lattice QCD~\cite{PHMC_Takaishi_Forcrand,PHMC_JLQCD}. 

The Kogut-Susskind (KS) fermion action has also been widely used in full QCD 
simulations. 
For the KS action, the application of the HMC algorithm is restricted 
to a multiple of four flavors due to the four-fold Dirac fermion content of 
a single KS fermion. 
Even if one adopts the usual assumption that the $1/4$ power of 
the KS fermion determinant provides a lattice discretization 
of a single Dirac fermion determinant, 
efficient exact algorithms have not been known for two- or single-flavor quark with 
the KS formalism.
For this reason, dynamical KS fermion simulations for two-  or three-flavor 
QCD are made with the $R$-algorithm~\cite{HybridR} even today, 
which is an approximate algorithm. 

The $R$-algorithm involves a systematic error from a finite step size of 
the molecular dynamics integrator. 
Strictly speaking, a careful extrapolation of physical quantities 
to the zero step size is required.
Since this is too computer time consuming, numerical simulations are usually 
carried out at a finite step size which is chosen so that the systematic error 
is considered invisible compared to the statistical error. 
While checks on small lattices are usually made to ensure smallness of the 
systematic error at least for several quantities,  
the possibility that other physical quantities are spoiled by the finite step 
size effects is difficult to exclude.  Even such checks become progressively 
more difficult as smaller quark masses require vastly increasingly computer time. 
Clearly, an exact and efficient algorithm is desired for dynamical QCD 
simulations with the KS fermion action not only for two flavors but also 
for a single flavor of quarks. 

In this article, we propose an exact algorithm for KS fermions which is 
capable of simulating an arbitrary number of flavors.  Our algorithm is 
an application of the PHMC algorithm.  It is an extension of the idea of 
the Rational Hybrid Monte Carlo (RHMC) algorithm~\cite{RHMC_PHMC} 
put forward by Horv\'{a}th, Kennedy, and Sint a few years ago. 
They briefly described their idea and tested their algorithm together 
with the PHMC algorithm for two- and four-flavor QCD with the KS fermions on 
a small size lattices.  We shall comment on the difference between their 
algorithm and ours below. 
Recently Hasenfratz and Knechtli~\cite{Hasenfratz_Knechtli} also proposed 
an exact algorithm for KS fermions with hyper-cubic smeared links, 
which makes use of the polynomial approximation and global update 
algorithm. The algorithm is considered to be effective for the action for 
which the HMC type algorithm cannot be applied. 

The outline of our algorithms goes as follows. 
Applying a polynomial approximation to the fractional power of the 
KS fermion matrix, 
we rewrite the original partition function to a form suitable for 
the PHMC algorithm.
The resulting effective partition function has two parts; 
one part is described by an effective action for the polynomial approximation 
of the fermion action, which can be treated by the HMC algorithm. 
The other part is the correction term which removes the systematic errors
from the polynomial approximation. The correction term can be evaluated
by the Kennedy-Kuti~\cite{Kennedy_Kuti} noisy Metropolis test,
which has been successfully used in the multi-boson 
algorithm~\cite{Noisy_Metropolis}.
With this combination, we can make an exact algorithm for KS fermions.
In this work we describe the details of the algorithm, 
and report results of our numerical test on the applicability of the algorithm 
to realistic simulations.

Since the polynomial approximation to rewrite the partition function
is not unique, there can be several realizations of the PHMC algorithm 
for the KS fermion.
We construct two types of realizations of the polynomial 
approximation to the PHMC algorithm with $N_f$ quark flavors:
\begin{description}
\item[Case A] Use a polynomial which approximates the $N_f/8$ power of 
              the KS fermion matrix which corresponds to $N_f/2$ quark flavors.
              Introducing a single pseudo-fermion field with squaring 
              the fermion matrix, we obtain $N_f$ flavors of quarks.
\item[Case B] Use an even-order polynomial which approximates the $N_f/4$ 
              power of the KS fermion matrix, which corresponds to $N_f$ flavors of quarks.   
              To express $N_f$ quark flavor with a single pseudo-fermion field,
              the even-order polynomial is factored into a product of two polynomials 
              using the method by 
              Alexandrou \textit{et al.}~\cite{Nf_one_Alexandrou}.
\end{description}
We estimate the computational cost of the two algorithms in terms of 
the number of multiplication by the KS fermion matrix, and
find that the algorithm B is roughly a factor two more efficient. 

We investigate the property and efficiency of the Chebyshev polynomial 
approximation as an example of the polynomial approximation. 
The method to split the even-order polynomial into two polynomials, which is
used in case B, is also described.
The limitation of our method for splitting the even-order polynomial is discussed.

The Kennedy-Kuti noisy Metropolis test involves a non-trivial factor. 
Since we take the fractional power of the KS operator, the correction term also
includes fractional powers of fermion matrices. 
In order to evaluate the measure for the noisy Metropolis test acceptance rate, 
we need to evaluate the fractional power of the correction matrix.
To do this, we make use of the Lanczos-based Krylov subspace method by 
Bori\c{c}i~\cite{Borici_matrix_function}, originally proposed for the 
inverses square root of the squared Hermitian
Wilson-Dirac operator in the Neuberger's overlap fermion. 
We modify the Bori\c{c}i's algorithm to our purpose.

Finally we investigate the validity and property of 
the PHMC algorithm (case B) in the case of two flavors of quarks numerically.
We first confirm that our algorithm works correctly using a small lattice of a 
size $8^3\times 4$, where a comparison to the $R$-algorithm is performed.
On an $8^3\times 16$ lattice we investigate the mass dependence of the 
algorithm.
We find that for quark masses lighter than 
$m_{\mathrm{PS}}/m_{\mathrm{V}}\sim 0.60$
the polynomial with order larger than $O(600)$ is required. 
Finally, we apply our algorithm to a moderately large lattice of $16^4$ with a
rather heavy quark mass $m_{\mathrm{PS}}/m_{\mathrm{V}}\sim 0.69$ as a
prototype for future realistic simulations.
On this lattice violation of reversibility and convergence of the 
Lanczos-based Krylov subspace method for the noisy Metropolis test are investigated. 
We find satisfactory results; 
there is no visible reversibility violation, and 
the Krylov subspace method converges within the limit of double precision 
arithmetic.  The test run on the large size lattice
shows reasonable efficiency on the computational time. 

The rest of the paper is organized as follows.
In section~\ref{sec:sec2}, we introduce the lattice QCD partition
function with the KS fermions, 
and rewrite it to a form suitable for the PHMC algorithm.
We give two forms for the partition function as described above. 
The outline of the PHMC algorithm is also shown in this section.
In section~\ref{sec:sec3}, we describe the Chebyshev polynomial as a specific
choice for the polynomial approximation. 
The error of the polynomial approximation is investigated.
The molecular dynamics (MD) step with the polynomial approximation  
in the PHMC algorithm is explained in section~\ref{sec:sec4}. 
The details of the noisy Metropolis test is given in section~\ref{sec:sec5}.
We estimate the computational cost of the algorithms in section~\ref{sec:sec6},
where we find that the case B algorithm is better.
In section~\ref{sec:sec7}, we show the test results with the PHMC algorithm.
Conclusions are given in the last section.

\section{Effective Action for PHMC algorithm}
\label{sec:sec2}

The QCD partition function for $N_f$ flavors of quarks using the 
KS fermion is defined by
\begin{equation}
  Z=\int{\cal D}U \det[D]^{N_{f}/4} e^{-S_{g}[U]},
\end{equation}
where $S_{g}[U]$ is the gauge action, $U_{\mu}(n)$ is gauge links, 
and $\det[D]$ is the determinant of the KS fermion operator $D$. 
The KS fermion operator $D$ is defined by 
\begin{equation}
  D(n,m) = am \delta_{n,m} + \frac{1}{2}\sum_{\mu}\eta_{\mu}(n)
  \left[ U_{\mu}(n)\delta_{n+\hat{\mu},m}
       - U^{\dag}_{\mu}(n-\hat{\mu})\delta_{n-\hat{\mu},m},
  \right]
\label{eq:Staggered_Op}. 
\end{equation}
where $am$ is the bare quark mass with lattice unit $a$, 
and $\eta_{\mu}(n)$ is the usual KS fermion phase.
In our work we adopt the usual assumption that taking the fourth root of 
the KS fermion operator represents a lattice discretization of a single 
Dirac fermion operator in the continuum. 

The determinant of the KS fermion operator can be rewritten as
\begin{equation}
  \det[D] = \det\left[
                  \begin{array}{cc}
                      am \mathbf{1}_{ee}  &   M_{eo} \\
                       M_{oe}  &  am \mathbf{1}_{oo}
                  \end{array}
               \right]
   =  \det\left[
                  \begin{array}{cc}
                         \mathbf{1}_{ee}  &   0  \\
                             0  &  \hatD
                  \end{array}
               \right],
\end{equation}
where $\hatD = (am)^{2} \mathbf{1}_{oo} - M_{oe}M_{eo}$ with
 $M_{eo} (M_{oe})$ the hopping matrix from odd site to even site 
(even site to odd site) defined in Eq.~(\ref{eq:Staggered_Op}). 
Since $\hatD$ is nothing but the odd part of $D^{\dag} D$,
the eigenvalues are real and positive semi-definite, which enable us
to take the fractional power of the KS fermion operator except for
vanishing quark masses.
Thus the QCD partition function is reduced to
\begin{equation}
  Z=\int {\cal D}U \det[\hatD]^{N_{f}/4} e^{-S_{g}[U]}.
\label{eq:QCD_PF}
\end{equation}

To apply the PHMC algorithm, we approximate the fractional power of 
$\hatD$ by a polynomial of $\hatD$. 
We consider two methods for the polynomial approximation.  
We restrict ourselves to the case that the number of flavors is one or two. 
Generalization to any integer flavors is achieved by combining the single-flavor 
and two-flavor cases.

\paragraph*{Case A}

We introduce a polynomial $P_{\Npoly}[x]$ of order 
$\Npoly$, which approximates $x^{-N_{f}/8}$
for real and positive (non-zero) $x$ as
\begin{equation}
  x^{-N_{f}/8}\sim P_{\Npoly}[x]=\sum_{i=0}^{\Npoly} c_{i}x^{i},
\label{eq:DefPoly}
\end{equation}
where $c_{i}$'s are real coefficients.
The property that the eigenvalues of the KS fermion operator
$\hatD$ are bounded below by $(am)^2$ enables us to substitute 
$\hatD$ into Eq.~(\ref{eq:DefPoly})
to approximate the fractional power of the KS fermion operator.
Using this polynomial, we can rewrite the determinant as
\begin{eqnarray}
\det\left[\hatD\right]^{N_{f}/4}&=&\left(
\frac{\det\left[\hatD\left(P_{\Npoly}[\hatD]\right)^{8/N_{f}}\right]}
{\det\left[\left(P_{\Npoly}[\hatD]\right)^{8/N_{f}}\right]}
\right)^{N_{f}/4} \nonumber \\
&=&\frac{\det\left[W[\hatD]\right]^{N_{f}/4}}
{\det\left[P_{\Npoly}[\hatD]\right]^2},
\label{eq:case1det}
\end{eqnarray}
where we introduced 
\begin{equation}
W[\hatD]=\hatD\left(P_{\Npoly}[\hatD]\right)^{8/N_{f}},
\label{eq:CorrCaseA}
\end{equation}
whose deviation from the identity matrix indicates the residual of 
the polynomial approximation.
We refer to $W[\hatD]$ as the correction matrix. Note that 
the exponent of $P_{\Npoly}$ becomes an integer when $N_f=2$ or $N_f=1$
as we assumed.
Introducing pseudo-fermion fields to the denominator of Eq.~(\ref{eq:case1det}),
we obtain
\begin{eqnarray}
\det\left[\hatD\right]^{N_{f}/4}&=&
\det\left[W[\hatD]\right]^{N_{f}/4}
\int{\cal D}\phi_{o}^{\dag}{\cal D}\phi_{o} e^{-S_{q}[U,\phi_{o}^{\dag},\phi_{o}]},\\
S_{q}[U,\phi_{o}^{\dag},\phi_{o}]&=&
\left|P_{\Npoly}[\hatD]\phi_{o}\right|^{2}.
\label{eq:Quark_Action_case1}
\end{eqnarray}
Thus the QCD partition function Eq.~(\ref{eq:QCD_PF}) becomes
\begin{equation}
  Z=\int {\cal D}U{\cal D}\phi_{o}^{\dag}{\cal D}\phi_{o}
    \det[W[\hatD]]^{N_{f}/4} e^{-S_{g}[U]-S_{q}[U,\phi_{o}^{\dag},\phi_{o}]}.
\label{eq:Effective_Action1}
\end{equation}

\paragraph*{Case B}

We define a polynomial $P_{\Npoly}[x]$ with $\Npoly$ even 
to approximate $x^{-N_{f}/4}$ for real and positive (non-zero) $x$
by
\begin{equation}
  x^{-N_{f}/4}\sim P_{\Npoly}[x]=\sum_{i=0}^{\Npoly} c_{i}x^{i}.
\label{eq:DefPolyB}
\end{equation}
Similarly to the case A, we can rewrite the determinant as
\begin{eqnarray}
\det\left[\hatD\right]^{N_{f}/4}&=&\left(
\frac{\det\left[\hatD\left(P_{\Npoly}[\hatD]\right)^{4/N_{f}}\right]}
{\det\left[\left(P_{\Npoly}[\hatD]\right)^{4/N_{f}}\right]}
\right)^{N_{f}/4}\nonumber \\
&=&\frac{\det\left[W[\hatD]\right]^{N_{f}/4}}
{\det\left[P_{\Npoly}[\hatD]\right]},
\label{eq:case2detprot}
\end{eqnarray}
where 
\begin{equation}
W[\hatD]=\hatD\left(P_{\Npoly}[\hat{D}_{oo}]\right)^{4/N_{f}}.  
\label{eq:CorrCaseB}
\end{equation}
For $\Npoly$ even, $P_{\Npoly}[x]$ can be factored into two 
polynomials as employed in the multi-boson algorithm for single-flavor 
QCD~\cite{Nf_one_Alexandrou}.
Making use of this property, we obtain
\begin{equation}
\det\left[\hatD\right]^{\frac{N_f}{4}}
=\frac{\det\left[W[\hatD]\right]^{N_{f}/4}}
{\left|\det\left[Q_{\Npoly}[\hatD]\right]\right|^{2}},
\label{eq:case2det}
\end{equation}
where $Q_{\Npoly}[x]$ is defined by
\begin{equation}
\left|Q_{\Npoly}[x]\right|^{2}=P_{\Npoly}[x],\ \ \ \ 
 Q_{\Npoly}[x]=\sum^{\Npoly/2}_{i=0}d_i x^i,
\label{eq:DefQpoly1}
\end{equation}
with complex coefficients $d_i$. The factoring is not unique. 
In the next section we describe the method for dividing the polynomial.
Introducing pseudo-fermion fields to the denominator of Eq.~(\ref{eq:case2det}),
we obtain the QCD partition function for the PHMC algorithm in the case B as
\begin{eqnarray}
  Z&=&\int {\cal D}U{\cal D}\phi_{o}^{\dag}{\cal D}\phi_{o}
    \det[W[\hatD]]^{N_{f}/4} e^{-S_{g}[U]-S_{q}[U,\phi_{o}^{\dag},\phi_{o}]}, 
\label{eq:Effective_Action2} \\
S_{q}[U,\phi_{o}^{\dag},\phi_{o}]&=&
\left|Q_{\Npoly}[\hatD]\phi_{o}\right|^{2}.
\label{eq:Quark_Action_case2}
\end{eqnarray}

The PHMC algorithm follows two steps:
\begin{enumerate}
\item The HMC step with the effective action 
 Eq.~(\ref{eq:Effective_Action1}) for the case A 
(Eq.~(\ref{eq:Effective_Action2}) for the case B).
\item The noisy Metropolis test to remove the systematic error 
 represented by $W[\hatD]$ in Eq.~(\ref{eq:case1det}) for the case A 
(Eq.~(\ref{eq:case2det}) for the case B).
\end{enumerate}
The noisy Metropolis test in step (2) has been used in the multi-boson 
algorithm~\cite{Noisy_Metropolis}. 
The reweighting technique~\cite{PHMC_Frezzotti_Jansen} 
and stochastic noisy estimator~\cite{PHMC_Takaishi_Forcrand} 
method can be applied to remove the systematic error. 
In this paper we employ the noisy Metropolis test
to make our algorithm exact~\cite{Noisy_Metropolis,JLQCD}.

The original idea by Horv\'{a}th \textit{et al.} differs from ours.
We separate the action into two pieces; the effective action with 
polynomial approximation and the correction factor.
They do not separate the full action and
apply the Kennedy-Kuti noisy estimator for the energy conservation 
violation $dH$ itself to make the algorithm exact, 
while we apply the method to the correction factor.
It is not known which approach is more efficient.
We leave a study of this issue to future work as a comparison of the two 
algorithms is beyond the scope of the present paper. 

\section{Polynomial approximation}
\label{sec:sec3}

We employ the Chebyshev polynomial approximation to $x^{-s}$
with a real positive $x$, where $s$ takes the following values: 
$1/2$, $1/4$, $1/8$,
depending on the choice of case A or B, and on $N_f$. 
We explain the application to the KS fermion operator and 
investigate the relation among the order of polynomial, the residual of 
the approximation, and the choice of case A or B.

Several polynomial approximations for $1/x$ have been proposed in
studies of the multi-boson algorithm. They include Chebyshev~\cite{Luscher_MB},
adapted~\cite{Adopted}, and least square~\cite{Montvay_LS} polynomials.
The choice of the polynomial affects the efficiency, {\it i.e.,} 
how much one can decrease the polynomial order so as to make
the correction matrix as close to the unit matrix as possible.
Since this is not a problem of the simulation algorithm itself,  
we employ the simple choice of the Chebyshev polynomial approximation for
the fractional power of the KS fermion 
operator.

\subsection{Chebyshev approximation}

The Chebyshev polynomial expansion of $x^{-s}$ is
\begin{equation}
 x^{-s}=\left[1+(1-\epsilon)y\right]^{-s}
 = \sum_{k=0}^{\infty}c_{k}T_{k}[y],
\label{eq:ExactChebyshevExpansion}
\end{equation}
where $y=(x-1)/(1-\epsilon)$, $\epsilon$ is optimized so as to
satisfy $y \in [-1,1]$ depending on the support of $x$ and
$\epsilon \in (0,1)$. We restrict the support of exponent $s$ to $s\in (0,1]$.
~\footnote{The fractional power inverse of a matrix is also given by
           Gegenbauer polynomial expansion~\cite{Bunk}.}
$T_{k}[x]$ is the $k$-th order Chebyshev polynomial defined by 
\begin{equation}
T_{k}[x] = \cos(k \arccos(x)).
\end{equation}
The polynomial has the following recurrence formula:
\begin{equation}
  T_{k}[x]= \alpha_{k-1} x T_{k-1}[x] + \beta_{k-2}T_{k-2}[x]\ \ \  (k \ge 1),
\label{eq:Chebyshev_Reccurence}
\end{equation}
where $\alpha_{k}=2/(1+\delta_{k,0})$ ($k \ge 0$), 
$\beta_{-1}=0$, $\beta_{k}=-1$ ($k \ge 0$), and $T_{0}[x]=1$.
The Chebyshev polynomial expansion coefficients $c_{k}$ 
in Eq.~(\ref{eq:ExactChebyshevExpansion}) are calculated as
\begin{eqnarray}
  c_{k}&=&{\int_{-1}}^{1}[1+(1-\epsilon)y]^{-s}\frac{T_{k}[y]}{\sqrt{1-y^2}}dy\nonumber\\
       &=&\frac{2 r^k }{1+\delta_{k,0}} (1+r^2)^{s} 
          F(s,s+k;1+k;r^2)\frac{\Gamma(s+k)}{\Gamma(s)\Gamma(1+k)},
  \label{eq:ExactPolyCoef}  
\end{eqnarray}
where $r=\left(-1+\sqrt{\epsilon(2-\epsilon)}\right)/(1-\epsilon)$, 
$F(\alpha,\beta;\gamma;z)$  Gaussian hyper-geometric function, and
$\Gamma(z)$ Gamma function.
Truncating Eq.~(\ref{eq:ExactChebyshevExpansion}) at order $\Npoly$,
we approximate $x^{-s}$ by 
\begin{equation}
x^{-s}=\left[1+(1-\epsilon)y\right]^{-s}
\sim P_{\Npoly}[x] = \sum_{k=0}^{\Npoly}c_{k}T_{k}[y].
\label{eq:ChevApprox}
\end{equation}
The truncation error is bounded by
\begin{equation}
\label{eq:ErrorBound}
  \left|x^{-s}-P_{\Npoly}[x]\right| \le 
\frac{2}{\Gamma(s)} \left(\frac{1+r^2}{1-r^2}\right)^{s} \frac{(-r)^{\Npoly+1}}{1+r}, 
\end{equation}
where $r$ takes a value in the region $-1\le r\le 0$.  
This inequality is not optimal.  It demonstrates, 
however, an exponential decrease of the residual with increasing $\Npoly$.
When $\epsilon\ll 1$, it is expected that $\Npoly \propto \sqrt{\epsilon}$
at a constant residual.
        
The operator corresponding to $y$ in the above relation is given
by shifting and changing the normalization of the KS fermion operator:
\begin{equation}
\hat{D'}_{oo}= \mathbf{1}_{oo} - (a\lambda)^{2}\hat{M}_{oo} = 
\frac{2}{2 (am)^2 +(a\Lambda_{\mathrm{max}})^{2}} \hatD,
\label{eq:KS_OP}
\end{equation}
where 
$(a\lambda)^{2}=2(a\Lambda_{\mathrm{max}})^{2}/(4 (am)^{2} + 2 (a\Lambda_{\mathrm{max}})^{2})$, 
and $\hat{M}_{oo}=\mathbf{1}_{oo}+2 M_{oe}M_{eo}/(a\Lambda_{\mathrm{max}})^{2}$. 
$a\Lambda_{\mathrm{max}}$ is introduced to keep all eigenvalues of 
$-\hat{M}_{oo}$ in the domain $[-1,1]$.
Since the normalization in front of $\hatD$ in Eq.~(\ref{eq:KS_OP}) can be 
absorbed into the normalization of the partition function, 
we use $\hat{D'}_{oo}$ instead of $\hatD$ and omit the prime symbol 
from $\hat{D'}_{oo}$ in the rest of the paper.
For the free case, $a\Lambda_{\mathrm{max}}=2$ 
is sufficient to satisfy the condition for the eigenvalues of $-\hat{M}_{oo}$.
In the interacting case, it becomes larger than two. 
This is seen, for example, in a study for SU(2) lattice gauge theory where 
the complete eigenvalue distribution has been 
investigated \cite{Kalkreuter_Eigen_Dist}.
With this expression for the KS fermion operator, the polynomial approximation
of $\hatD^{-s}$ becomes
\begin{equation}
  \hatD^{-s} \sim
  P_{\Npoly}[\hatD]=\sum_{i=0}^{\Npoly}c_{k}T_{k}[-\hat{M}_{oo}],
\label{eq:PolyCheb}
\end{equation}
where $c_{i}$ is obtained from Eq.~(\ref{eq:ExactPolyCoef}) 
with $\epsilon=1-(a\lambda)^{2}$.
The exponent $s$ is chosen to be $s=N_{f}/8$ for the case A and 
$s=N_{f}/4$ for the case B.

For the case B, we have to solve Eq.~(\ref{eq:DefQpoly1}) to obtain 
the half-order polynomial $Q_{\Npoly}$. 
Here we choose to construct $Q_{\Npoly}$ so as to have the form of 
the Chebyshev polynomial expansion as Eq.~(\ref{eq:PolyCheb}). 
Since this problem is rather complicated, 
we postpone the discussion to the subsection at the end of this section, 
and proceed assuming that the polynomial $Q_{\Npoly}$ and its coefficients 
$d_{i}$ are already given. 

Given the Chebyshev polynomial expansion coefficients $c_{m}$ ($d_{m}$),
we can evaluate the polynomial $P_{\Npoly}[x]$ ($Q_{\Npoly}[x]$)
using the Clenshaw's recurrence formula~(for example, see~\cite{NumericalRecipes}). 
When $x$ is the KS fermion operator $\hatD$, the multiplication of the 
operator $P_{\Npoly}[\hatD]$ on a vector $v_{o}$ is carried out 
by the following three step recurrence formula:
\begin{equation}
  y^{(k)}_{o}=\alpha_{k}(-\hat{M}_{oo})y^{(k+1)}_{o} 
             +\beta_{k} y^{(k+2)}_{o} + c_{k} v_{o},
\label{eq:PolyRecRel}
\end{equation}
where $y^{(k)}$ is a working vector labeled $k$, 
$\alpha_{k}$ and $\beta_{k}$ are
given in Eq.~(\ref{eq:Chebyshev_Reccurence}), 
and $c_{k}$ is the Chebyshev polynomial 
expansion coefficient.
Solving for $y^{(k)}_{o}$ with this equation 
from $k=\Npoly$ to $k=0$ with the initial condition
$y^{(\Npoly)}_{o}=y^{(\Npoly+1)}_{o}=0$,  
we obtain
\begin{equation}
  P_{\Npoly}[\hatD] v_{o}= y^{(0)}_{o}.
\end{equation}
For $Q_{\Npoly}[\hatD]v_{o}$, 
$d_{k}$ is used instead of $c_{k}$ and $k$ runs from $\Npoly/2$ to $0$.
Note that we need not store all working vectors $y^{(k)}_{o}$;  
only two working vectors are required for the computation.
This method can be applied to any matrix polynomial which has the 
same structure for the recurrence relation as 
Eq.~(\ref{eq:Chebyshev_Reccurence}). 
The computational cost to calculate $P_{\Npoly}[\hatD]v_{o}$ ($Q_{\Npoly}[\hatD]v_{o}$)
is $\Npoly$ ($\Npoly/2$) by means of the number of
matrix-vector multiplication.

\begin{figure}[t]
\begin{center}
  \includegraphics[scale=\figscale]{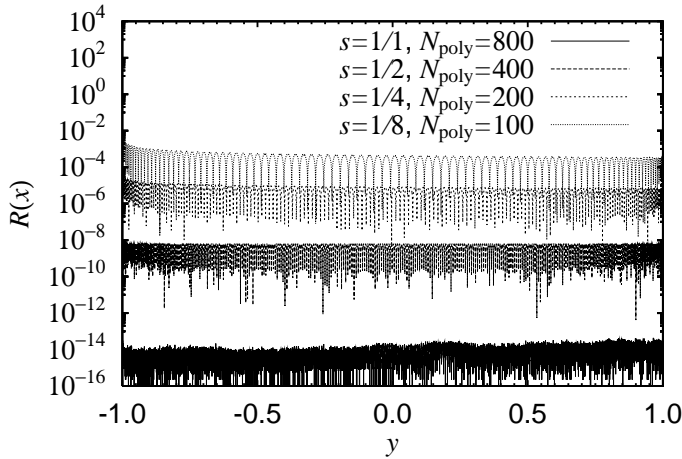}
  \caption{\label{fig:Poly_Res}  
      $R(x)$ as a function of
      $y=(x-1)/(1-\epsilon)$ at $\Npoly/s=800, \epsilon = 1/1000$.}
\end{center}
\end{figure}
\begin{figure}[t]
\begin{center}
  \includegraphics[scale=\figscale]{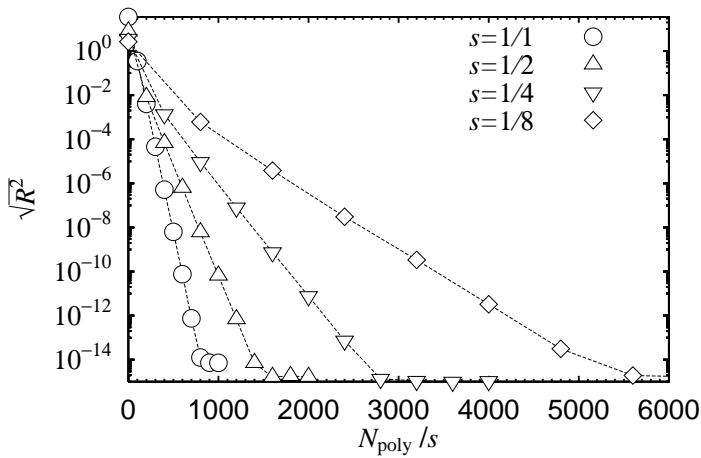}
  \caption{\label{fig:Residual}
           $\Npoly/s$ dependence of the integrated residuals $\sqrt{R^2}$.
           $\epsilon=1/1000$ are plotted.}
\end{center}
\end{figure}

Now we discuss the quality of the polynomial approximation.
We evaluate the polynomial approximation residual using
\begin{equation}
R(x)=\left|x (P_{\Npoly}[x])^{1/s}-1\right|.
\label{eq:Res}
\end{equation}
Figure~\ref{fig:Poly_Res} shows the residual of the approximation
as a function of $y=(x-1)/(1-\epsilon)$. 
The calculation is performed using Clenshaw's recurrence in double 
precision arithmetic.
In order to compare the approximation at the fixed computational cost,
the number of matrix-vector multiplication, in calculating the correction 
matrix irrespective of the choice of $s$ (case A or B and $N_f$), 
we keep $\Npoly/s$ constant ($\Npoly/s=800$) as an example.
We observe that the approximation becomes worse for 
smaller $s$ ($s=1$ case reaches the limit of double precision arithmetic).
We also investigate the $\Npoly/s$ dependence of 
the polynomial approximation.
In figure~\ref{fig:Residual} we plot the $\Npoly/s$ dependence 
of the residual defined by
\begin{equation}
  \label{eq:Residual}
  \sqrt{R^{2}}=\sqrt{{\int_{-1}}^{1} dy R(x(y))^{2}}.
\end{equation}
Clear exponential decay is observed. 
The dependence of the slope on the choice of $s$ indicates
that the computational cost increases with decreasing $s$.
In our case, defining the polynomial order by $\Npoly^{A}$ and $\Npoly^{B}$ 
for the case A and B, respectively, it is expected that
$\Npoly^{A}/(N_f/8) \sim 2 \Npoly^{B}/(N_f/4)$ holds when
we require that the integrated residual takes a similar value for the two cases.
We thus find $\Npoly^{A} \sim \Npoly^{B}$ at the same value of the 
polynomial residual.

\subsection{Determination of $Q_{\Npoly}$ in case B}
\label{subsec:subsec}

Here we discuss how to solve Eq.~(\ref{eq:DefQpoly1}) to obtain 
the half-order polynomial $Q_{\Npoly}$. 
In our work $Q_{\Npoly}$ should have the Chebyshev polynomial expansion 
form of Eq.~(\ref{eq:PolyCheb}).
A simple procedure to obtain $Q_{\Npoly}$ is as follows:  
(i) express Eq.~(\ref{eq:ChevApprox}) as a polynomial of $y$ 
instead of the expansion of the Chebyshev polynomial, 
(ii) split the polynomial into the product of two polynomials like 
Eq.~(\ref{eq:DefQpoly1}) by means of usual polynomial expansion, 
and (iii) recover the Chebyshev polynomial expansion 
for the half-order polynomial. 
However, this method has a numerical problem in step (iii).
In order to make this point clear, and to present an alternative procedure,  
let us elaborate on the procedure above. 

In the step (i), we expand the Chebyshev polynomial to write 
\begin{equation}
 P_{\Npoly}[x]=
 \sum_{k=0}^{\Npoly}c_{k}T_{k}[y]= \sum_{k=0}^{\Npoly} c'_{k} y^{k},
\label{eq:ChevToPoly}
\end{equation}
where $c'_{k}$ can be obtained from the original $c_{k}$ using 
the appropriate recurrence formula (see \textit{ex.}~\cite{NumericalRecipes}). 
In the step (ii), finding the roots of the polynomial 
$\sum_{k=0}^{\Npoly} c'_{k} y^{k}=0$, we obtain the product representation:
\begin{equation}
 \sum_{k=0}^{\Npoly} c'_{k} y^{k}= c'_{\Npoly}\prod_{k=1}^{\Npoly}(y-z_{k}),
\end{equation}
where $z_{k}$'s are the roots of the polynomial. Because $\Npoly$ is even 
and the coefficients $c'_{k}$'s are real, the root $z_{k}$ pairs with 
its complex conjugate $z_{k'}=z^{*}_{k}$.
Thus we can split the polynomial to the product of two polynomials~\cite{Nf_one_Alexandrou}: 
\begin{eqnarray}
\sum_{k=0}^{\Npoly} c'_{k} y^{k}&=&
c'_{\Npoly}\prod_{k=1}^{\Npoly/2}(y-z_{j(k)})(y-z^{*}_{j(k)}),\nonumber\\
&=&
\left|\left(c'_{\Npoly}\right)^{1/2}\prod_{k=1}^{\Npoly/2}(y-z_{j(k)})\right|^{2},
\end{eqnarray}
where $j(k)$ is a reordering index to represent the pairing condition above.
The explicit form of the reordering is not unique 
depending on how to distribute the complex pairs in the monomials, 
and several methods have been proposed~\cite{Montvay_LS,Ordering}. 

In the third step (iii), we calculate the half-order polynomial 
according to 
\begin{eqnarray}
  \left(c'_{\Npoly}\right)^{1/2}\prod_{k=1}^{\Npoly/2}(y-z_{j(k)})&=&
     \sum_{k=0}^{\Npoly/2} d'_{k} y^{k}, \nonumber \\
&=&  \sum_{k=0}^{\Npoly/2} d_{k} T_{k}[y],
\end{eqnarray}
where $d'_{k}$'s are obtained from $z_{k}$ and $c'_{\Npoly}$ by expanding 
the product representation, and $d_{k}$'s are extracted from $d'_{k}$'s 
using an appropriate reverse recurrence formula~\cite{NumericalRecipes} (\textit{i.e.} the 
relation  opposite to Eq.~(\ref{eq:ChevToPoly})).
In this way, we could derive $d_{k}$'s  from $c_{k}$'s
so as to satisfy
\begin{equation}
 P_{\Npoly}[x]=\sum_{k=0}^{\Npoly}c_{k}T_{k}[y]= 
\left|\sum_{k=0}^{\Npoly/2} d_{k} T_{k}[y]\right|^2 =
\left|Q_{\Npoly}[x]\right|^{2}.
\label{eq:ChebSplit}
\end{equation}
Unfortunately, we find that the reverse recurrence to extract 
the Chebyshev polynomial expansion coefficients $d_{k}$ from the 
usual polynomial coefficients $d'_{k}$'s is numerically unstable~\cite{NumericalRecipes}. 
Therefore we decided to directly solve Eq.~(\ref{eq:ChebSplit}) with 
respect to $d_{k}$.

Using the addition relation, $T_{k}[y]T_{l}[y]=(T_{k+l}[y]+T_{|k-l|}[y])/2$,
to Eq.~(\ref{eq:ChebSplit}),
we extract the following second-order simultaneous equations, 
\begin{equation}
  f_{k}\left(\{d\},\{d^{*}\}\right)=c_k, \ \ \ (k=0,\cdots,\Npoly),
\label{eq:SimEq}
\end{equation}
where $f_k\left(\{d\},\{d^{*}\}\right)$ depends on the sets
$\{d\}=\{d_1,d_2,\ldots,d_{\Npoly/2}\}$ and 
$\{d^{*}\}=\{d^{*}_1,d^{*}_2,\ldots,d^{*}_{\Npoly/2}\}$.
We do not write down the explicit form of $f_{k}\left(\{d\},\{d^{*}\}\right)$ 
here because of its length and complicated form.
The solution to these equations is not unique.  
This corresponds to the reordering ambiguity in the previous case of 
splitting the product representation.

\begin{figure}[t]
\begin{center}
  \includegraphics[scale=\figscale]{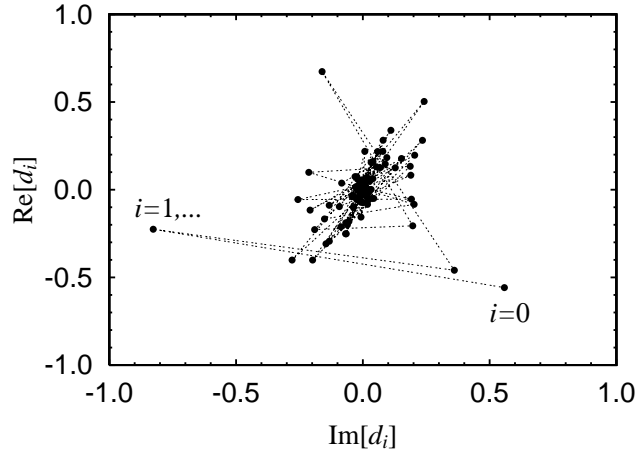}
  \caption{\label{fig:PolyQ_Coef}  
      Polynomial coefficients $d_{i}$ for $Q_{\Npoly}[x]$ 
      at $\epsilon=0.0001$, $\Npoly=600$, and $N_{f}=2$
      in complex plane.}
\end{center}
\end{figure}

\begin{figure}[t]
\begin{center}
  \includegraphics[scale=\figscale]{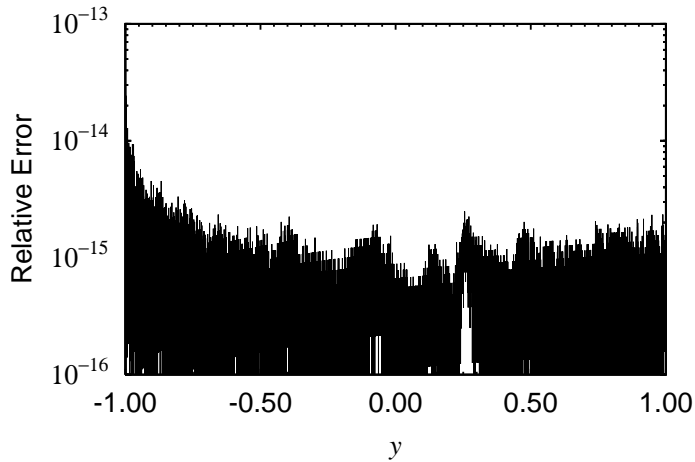}
  \caption{\label{fig:PolyQ_RelErr}  
      Relative error, 
      $|P_{\Npoly}[x]-Q_{\Npoly}[x]Q^{*}_{\Npoly}[x]|/|P_{\Npoly}[x]|$,
      plotted against $y$, where $x=1+(1-\epsilon)y$.
      Parameters are the same as those in Fig.~\ref{fig:PolyQ_Coef}.}
\end{center}
\end{figure}

We solve numerically the simultaneous equation Eq.~(\ref{eq:SimEq}) using
\textit{Mathematica} with desired accuracy starting from an initial
choice of $\{d\}$ and $\{d^{*}\}$. 
The accuracy of the solution is examined
by numerically evaluating the relation Eq.~(\ref{eq:ChebSplit}).
Figure~\ref{fig:PolyQ_Coef} shows the polynomial coefficients $d_{i}$
for $Q_{\Npoly}[x]$ derived by the direct method using Eq.~(\ref{eq:SimEq}). 
The accuracy of the solution stays at satisfactory level 
within double precision arithmetic as observed in 
Figure~\ref{fig:PolyQ_RelErr}, where the polynomials are evaluated with 
Clenshaw's recurrence formula in double precision arithmetic.


A practical limitation with our direct method is that it is rather slow. 
On a Linux PC with 1 GHz Pentium III CPU, we could find the coefficients 
only for $\Npoly < 800$ within a tolerable computational time.
A more efficient methods to solve Eq.~(\ref{eq:DefQpoly1}) is desired.

\section{Calculation of the polynomial pseudo-fermion force in the Hybrid Monte Carlo algorithm}
\label{sec:sec4}

We apply the usual HMC algorithm to the partition 
function Eq.~(\ref{eq:Effective_Action1}) (case A) or  
Eq.~(\ref{eq:Effective_Action2}) (case B),  introducing a fictitious time
and canonical momenta $P_{\mu}(n)$ to the link variables $U_{\mu}(n)$.
A nontrivial task is the calculation of the molecular dynamics (MD) force 
from the quark action expressed by the polynomial approximation.
We utilize the Clenshaw's recurrence formula for this purpose. 
Here we first discuss the variation of the polynomial of 
a matrix $A$ for the general case, and then describe the force calculation 
in the HMC algorithm.  

Let $P_{N}[A]$ be a matrix polynomial of A with order $N$
described by
\begin{equation}
  P_{N}[A] = \sum_{i=0}^{N}c_{i} \Phi_{i}[A],
\label{eq:Matrix_Polynomial}
\end{equation}
where $\Phi_{i}[A]$ is defined to have the following 
recurrence relation:
\begin{equation}
  \Phi_{i}[A]=\alpha_{i-1} A \Phi_{i-1}[A] + \beta_{i-2} \Phi_{i-2}[A].
\label{eq:Reccurence_Formula}
\end{equation}
In most cases $\Phi_{0}[A]$ is a constant and set to be unity.
As in Eq.~(\ref{eq:PolyRecRel}), $P_{N}[A]$ is evaluated by the
Clenshaw's recurrence formula:
\begin{equation}
  Y^{(k)}=\alpha_{k} A\ Y^{(k+1)} + \beta_{k} Y^{(k+2)} + c_{k}\mathbf{1},
\label{eq:Reccurence_A}
\end{equation}
where $Y^{(k)}$'s are working matrices, $Y^{(N+1)}=Y^{(N)}=0$,
$k$ runs from $N$ to $0$, and $P_{N}[A]=Y^{(0)}$.
We take the variation of Eq.~(\ref{eq:Matrix_Polynomial}) 
with respect to $A$, and denote it by $\delta P_{N}[A]$:
\begin{equation}
 \delta P_{N}[A] = \sum_{i=0}^{N}\delta \Phi_{i}[A] c_{i}
\label{eq:Derivative_Matrix_Polynomial}.
\end{equation}
The variation $\delta \Phi_{i}[A]$ also has the recurrence formula obtained
by differentiating Eq.~(\ref{eq:Reccurence_Formula}):
\begin{equation}
  \delta \Phi_{i}[A]=\alpha_{i-1}\delta \Phi_{i-1}[A] A
                    + \beta_{i-2}\delta \Phi_{i-2}[A]
                    +\alpha_{i-1}\Phi_{i-1}[A]\delta A.
\label{eq:Reccurence_Formula_Deriv}
\end{equation}
Substituting 
Eq.~(\ref{eq:Reccurence_A}) and Eq.~(\ref{eq:Reccurence_Formula_Deriv}) 
into Eq.~(\ref{eq:Derivative_Matrix_Polynomial}),
we have
\begin{equation}
\delta P_{N}[A] = \sum_{i=1}^{N}\alpha_{i-1}\Phi_{i-1}[A]\delta A\ Y^{(i)},
\label{eq:Deriv_Matrix_Polynomial}
\end{equation}
where $\delta\Phi_{0}[A]=0$ and
a similar technique to the Clenshaw's formula is used.

Applying Eq.~(\ref{eq:Deriv_Matrix_Polynomial}) to our problem, 
we can evaluate the variation of the pseudo-fermion action
Eq.~(\ref{eq:Quark_Action_case1}) with respect to the infinitesimal 
change of link variables as
\begin{eqnarray}
\delta S_{q}&=&\delta \left|P_{\Npoly}[-\hat{M}_{oo}]\phi_{o}\right|^{2}\nonumber \\
&=&\sum_{i=1}^{\Npoly}
   \left(\alpha_{i-1}T_{i-1}[-\hat{M}_{oo}]y^{(0)}_{o}\right)^{\dag}
   \left(-\delta \hat{M}_{oo}\right) y^{(i)}_{o} + \mbox{h.c.},
\end{eqnarray}
where $\mbox{h.c.}$ means the Hermitian conjugate of the preceding term,  
and $y_{o}^{(i)}$ satisfies Eq.~(\ref{eq:PolyRecRel}) with $v_{o}=\phi_{o}$.
We used the Hermiticity of $\alpha_{k}$ and $P_{\Npoly}[-\hat{M}_{oo}]$, and 
$y^{(0)}_{o}=P_{\Npoly}[-\hat{M}_{oo}]\phi_{o}$ was applied in the last line. 
A more convenient form of $\delta S_q$ for practical calculations is given by 
\begin{eqnarray}
\delta S_{q}
= - \frac{2}{(a\Lambda_{\mathrm{max}})^{2}}
    \sum_{i=1}^{\Npoly} \alpha_{i-1}
      \left.X^{(i)}\right.^{\dag} \delta M\ Z^{(i)} + \mbox{h.c.},
\label{eq:Force}
\end{eqnarray}
where $X^{(i)}$, $Y^{(i)}$, and $M$ are defined as
\begin{eqnarray}
  X^{(i)}&=& \left(
               \begin{array}{c}
                  -M_{eo} x_{o}^{(i)}\\
                          x_{o}^{(i)}
               \end{array}
             \right),\nonumber\\
  Z^{(i)}&=&\left(
               \begin{array}{c}
                   M_{eo} y_{o}^{(i)}\\
                          y_{o}^{(i)}
               \end{array}
             \right),\nonumber\\
\delta M&=&\left(
               \begin{array}{cc}
                    0          & \delta M_{eo} \\
                 \delta M_{oe} &    0 
               \end{array}
             \right),
  \label{eq:Memory_Force}
\end{eqnarray}
with
\begin{eqnarray}
  y_{o}^{(i)}&=&\alpha_{i}(-\hat{M}_{oo})y_{o}^{(i+1)}
               +\beta_{i}y_{o}^{(i+2)} + c_{i} \phi_{o}
   \ \ \ \ (i=\Npoly, \cdots, 0),\nonumber \\
  x_{o}^{(i)}&=&\alpha_{i-2}(-\hat{M}_{oo}) x_{o}^{(i-1)} +\beta_{i-3} x_{o}^{(i-2)}
   \ \ \ \ (i=2, \cdots, \Npoly),\nonumber\\
  x_{o}^{(1)}&=& y_{o}^{(0)}.
  \label{eq:Recurrence_Force}
\end{eqnarray}
In the actual calculation, 
we first calculate $y^{(i)}_{o}$ from $i=\Npoly$ to $i=0$
and store $Z^{(i)}$ on memory.  We then sum up each of the force contribution
$\left.X^{(i)}\right.^{\dag} \delta M \ Z^{(i)}$ evaluating 
$X^{(i)}$ from $i=1$ to $i=\Npoly$.  
A similar method, 
which has a more complicated form, has been obtained for the HMC 
algorithm with the overlap fermions by C.~Liu~\cite{LIU_OV_HMC}.
Since the even component of $X^{(i)}$ and $Z^{(i)}$ appear as a 
byproduct of the multiplication of $\hat{M}_{oo}$ in the recurrence 
equation, we do not need extra calculations for the even components.
The explicit form of the force contribution is obtained from these 
equations as usual.

We have described the force calculation for the case A as an example.
The force calculation for the case B is almost identical except for 
the replacement $\Npoly \to \Npoly/2$ and $c_{i} \to d_{i}$.

Eqs.~(\ref{eq:Force}), (\ref{eq:Memory_Force}), and (\ref{eq:Recurrence_Force})
do not contain an iterative procedure. 
This leads us to expect that the reversibility violation of 
the MD evolution is smaller than in the usual HMC algorithm with four-flavor 
KS fermions in which an iterative solver such as the Conjugate 
Gradient (CG) method is used to invert the KS fermion operator. 
Our implementation, however, still involves the possibility that 
round-off errors grow to violate the reversibility
in the summation of 
$\left.X^{(i)}\right.^{\dag} \delta M \ Z^{(i)}$ from $i=1$ to 
$i=\Npoly$.
We investigate this issue in 
section~\ref{sec:sec7} on a realistic size lattice.

The external field $\phi_{o}$ is generated at the beginning 
of every MD trajectory according to the pseudo-fermion action 
 Eq.~(\ref{eq:Quark_Action_case1}) (case A) 
(Eq.~(\ref{eq:Quark_Action_case2}) (case B)) using the 
global heat-bath method. This is achieved by solving the following 
equations with respect to $\phi_{o}$:
\begin{eqnarray}
  P_{\Npoly}[\hatD]\phi_{o}&=&\chi_{o} \mbox{\ \ \ \hfill(case A)},\\
  Q_{\Npoly}[\hatD]\phi_{o}&=&\chi_{o} \mbox{\ \ \ \hfill(case B)},
\end{eqnarray}
where $\chi_{o}$ is a Gaussian noise vector with unit variance.
Using the identity
\begin{eqnarray}
  \phi_{o}&=&(P_{\Npoly}[\hatD])^{-1}\chi_{o}\nonumber \\
&=&\hatD (P_{\Npoly}[\hatD])^{(8/N_f-1)}W[\hatD]^{-1}\chi_{o} \mbox{\ \ \ \hfill(case A)},
\label{eq:PseudoFieldA}\\
  \phi_{o}&=&(Q_{\Npoly}[\hatD])^{-1}\chi_{o}\nonumber\\
&=&\hatD (Q_{\Npoly}[\hatD])^{\dag}(P_{\Npoly}[\hatD])^{(4/N_f-1)}
W[\hatD]^{-1}\chi_{o} \mbox{\ \ \ \hfill(case B)},
\label{eq:PseudoFieldB}
\end{eqnarray}
we invert the coefficient matrix $P_{\Npoly}[\hatD]$ (case A)
($Q_{\Npoly}[\hatD]$ (case B)) using the CG solver to
$W[\hatD]$.

\section{Noisy Metropolis test}
\label{sec:sec5}

In order to make algorithm exact, we have to take into account the 
correction term $\det[W[\hatD]]^{N_{f}/4}$.
This is achieved by a noisy Kennedy-Kuti~\cite{Kennedy_Kuti} Metropolis test.
This method has been used in the multi-boson algorithm~\cite{Noisy_Metropolis}. 
In this section we explain the details of the noisy Metropolis test.

\subsection{Definition}

When a trial configuration $U'$ which is generated by the molecular dynamics
step in the HMC part of the algorithm is accepted at the HMC Metropolis test,
we make the noisy Metropolis test for the correction factor 
$\det[W[\hatD]]^{N_{f}/4}$.
The acceptance probability of the trial configuration $U'$ from an 
initial configuration $U$ is defined by
\begin{equation}
  P_{\mathit{corr}}[U\rightarrow U']=\min\left[1,e^{-dS[U,U']}\right],
\end{equation}
with 
\begin{equation}
  dS[U,U']=\left|\left.W[\hatD']\right.^{-N_{f}/8} 
  W[\hatD]^{N_{f}/8}\eta_{o}\right|^{2} - |\eta_{o}|^2,
\label{eq:dS_corr}
\end{equation}
where $\eta_{o}$ is a Gaussian random vector with unit variance.
This probability satisfies the detailed balance relation~\cite{Noisy_Metropolis}.
The factor $W[\hatD']$ is calculated on the trial configuration $U'$, while
$W[\hatD]$ on the initial configuration $U$.

Eq.~(\ref{eq:dS_corr}) can be modified to
\begin{eqnarray}
  dS[U,U']&=&\zeta_{o}^{\dag}\left.W[\hatD']\right.^{-N_{f}/4}\zeta_{o} 
  - |\eta_{o}|^2,  \nonumber \\
  \zeta_{o}&=&W[\hatD]^{N_{f}/8}\eta_{o}. 
\label{eq:dS_corr_simple}
\end{eqnarray}
We employ Eq.~(\ref{eq:dS_corr_simple}) for $dS$ in the present work. 
It is not known at present which of the expressions, 
Eq.~(\ref{eq:dS_corr}) or (\ref{eq:dS_corr_simple}),
is more useful for calculating $dS$ in respect of 
computational efficiency and accuracy, which is left for future studies. 

In order to evaluate Eq.~(\ref{eq:dS_corr_simple}),
we need to calculate the fractional power of the matrix $W[\hatD]$ (or $W[\hatD']$).
In Ref.~\cite{PHMC_JLQCD} we used a Taylor expansion method for 
the (inverse-)square root of the correction matrix with the $O(a)$-improved
Wilson fermions. 
In order to suppress the truncation error of the Taylor expansion
we explicitly calculate and monitor the residual for the Taylor expansion 
for which we need an extra computational cost.
The direct application of this method to Eq.~(\ref{eq:dS_corr}) requires
more computational overhead compared to the Wilson case,
since the residuals contains several application 
of $W[\hatD]^{N_{f}/8}$ and $W[\hatD]^{-N_{f}/4}$, 
(\textit{e.g.} for $N_{f}=2$ case, 8 times, and 4 times to define 
the residuals, respectively). 
Instead of this method we employ the Krylov subspace method 
to avoid the explicit calculation of the residuals as described
in the following.

Here we roughly estimate the order of polynomial $\Npoly$ required
to achieve a given acceptance rate at a volume $V$ and $a\Lambda_{\mathrm{max}}$.
The acceptance rate Eq.~(\ref{eq:dS_corr_simple}) behaves as 
$dS = \mbox{[constant]}\times V \times (-r)^{\Npoly+1}$, where $r$ is
defined in Eq.~(\ref{eq:ExactPolyCoef}).
To keep $dS < \delta$ with a small constant $\delta$ 
that maintains sufficient acceptance rate, we need
\begin{equation}
\Npoly > \frac{a\Lambda_{\mathrm{max}}}{2 am} 
\left(\ln V-\ln \delta +\mbox{[constant]}\right),
\end{equation}
where we used $r\sim -1 + 2 (am)/(a\Lambda_{\mathrm{max}})$ 
and $am\ll a\Lambda_{\mathrm{max}}$ from the definition of 
$r$ in Eq.~(\ref{eq:KS_OP}).
Therefore we need to increase $\Npoly$ linearly as increasing 
the condition number $(a\Lambda_{\mathrm{max}})/(am)$, and 
logarithmically with volume $V$.
A similar discussion on the required $\Npoly$ can be found in
Refs.~\cite{Noisy_Metropolis} in the literature of the multi-boson algorithm.

\subsection{The Krylov subspace method}

Since the matrix is Hermitian and positive definite with the KS fermion,
and already well preconditioned, we can take the fractional power with 
the Krylov subspace method with better efficiency~\cite{Borici_matrix_function,Bunk}.
A Lanczos-based Krylov subspace method was developed 
by Bori\c{c}i~\cite{Borici_matrix_function},
which was utilized for the calculation of inverse square root
of squared Hermitian Wilson-Dirac operator in the Neuberger's overlap operator.
The method is an application of the Krylov subspace method to calculate 
functions of large sparse matrices~\cite{Vorst}.
The Lanczos-based Krylov subspace method enables us to define 
a kind of residual without explicit residual calculation.
We apply this method to evaluate the fractional power of 
the correction matrix.

Consider a matrix function multiplied on a vector, $f(A)b$, 
with an $n\times n$ Hermitian matrix $A$ 
and an $n$ component vector $b$ in general. 
The $k$-dimensional orthogonal basis $Q_{k}=(q_{1},\cdots, q_{k})$, 
which spans the Krylov subspace with respect to the matrix $A$, is 
obtained by the Lanczos algorithm.
This basis satisfies the following condition:
\begin{equation}
  A Q_{k} = Q_{k} T_{k} + \beta_{k} q_{k+1} (e^{(k)}_{k})^{T},\ \ \
  q_{1}=\rho_{1}b,\ \ \  \rho_{1}=1/|b|,
\end{equation}
where $T_{k}$ is a $k\times k$ tridiagonal matrix
whose diagonal and sub-diagonal parts are 
$\alpha_{1}, \alpha_{2}, \cdots, \alpha_{k}$ and
$\beta_{1}, \beta_{2}, \cdots, \beta_{k-1}$, respectively.
$e^{(n)}_{m}$ is the unit vector in the $m$-th direction in $n$-dimension.
The superscript $T$ means transpose.

If $\beta_{k}$ is sufficiently small after $k$ iterations of 
the Lanczos method, the matrix function $f(A)$ acting on the 
vector $b$ is approximated by
\begin{equation}
  f(A)b \sim Q_{k}f(T_{k})e^{(k)}_{1}|b|.
\end{equation}
In our case, $A=W[\hatD]$, $f(x)=x^{-N_f/4}$, and $b=\eta_{o}$,
or $A=W'_{oo}$, $f(x)=x^{N_f/8}$, and $b=\zeta_{o}$.
The dimension $k$ is much smaller than that of $A$ when $A$ 
is close to unity.
Thus, the calculation of $f(A)$ is reduced to that of $f(T_{k})$ 
with smaller computational cost.

\begin{algorithm}[t]
\caption{Algorithm for matrix function $x=f(A)b$.}
\label{alg:alg1}
\begin{algorithmic}
  \STATE $\rho_{1}=1/|b|$; $q_{1}=b \rho_{1}$
  \FOR{$i=1,2,\cdots$}
    \STATE $r = A q_{i}$
    \IF{$i = 1$}
      \STATE $v=r$
    \ELSE
      \STATE $v=r-\beta_{i-1}q_{i-1}$
    \ENDIF
    \STATE $\alpha_{i}=q_{i}^{\dag} v$
    \STATE $v:=v-\alpha_{i} q_{i}$
    \FOR{$j=i,i-1,\cdots,2, 1$}
      \STATE $\gamma=q_{j}^{\dag}v$
      \STATE $v:=v-\gamma q_{j}$
    \ENDFOR
    \STATE $\beta_{i}=|v|$
    \IF{$i = 1$}
      \STATE $\rho_{i+1}=-{\rho_{i}\alpha_{i}}/{\beta_{i}}$
    \ELSE
      \STATE $\rho_{i+1}=-\left(\rho_{i}\alpha_{i}+\rho_{i-1}\beta_{i-1}\right)/{\beta_{i}}$
    \ENDIF
    \STATE $\mathit{err}_{1}=|1/\rho_{i+1}|$
    \STATE $q_{i+1}=v/\beta_{i}$
    \STATE set $(T_{i})_{i,i}=\alpha_{i}$, $(T_{i})_{i+1,i}=(T_{i})_{i,i+1}=\beta_{i}$, otherwise $(T_{i})_{i,j}=0$.
    \STATE diagonalize $T_{i}= U_{i}\Lambda_{i}U^{T}_{i}$, 
           where $T_{i}$ is the ($i\times i$) tridiagonal matrix.
    \STATE $t_{i}= U_{i}f(\Lambda_{i})U^{T}_{i}e^{(i)}_1/\rho_{1}$
    \FOR{$j=1,\cdots, i$}
      \STATE $x_{i}:=x_{i} + (t_{i})_{j} q_{j}$
    \ENDFOR
    \STATE ($\mathit{err}_{2}=|x_{i-1}-x_{i}|/|x_{i-1}|$)
    \IF{$\mathit{err}_{1} < \mathit{tol}$ ($\mathit{err}_{2} < \mathit{tol}$)}
      \STATE exit
    \ENDIF
  \ENDFOR
  \STATE solution $x=x_{i}\sim f(A)b$.
\end{algorithmic}
\end{algorithm}

Our algorithm is almost identical to that of Ref.~\cite{Borici_matrix_function}.
We show the algorithm to calculate the matrix function $f(A)b$
with the Lanczos-based method in Algorithm~\ref{alg:alg1}.
We compute $f(T_{k})$ by diagonalizing $T_{k}$ with LAPACK subroutines.
If the algorithm stops after $k$ iterations, we have an approximate solution
to the matrix function $f(A)$ given by $x_{k}\sim f(A)b$.

In our algorithm, a large cancellation error can occur 
in the Gram-Schmidt orthogonalization step 
because our correction matrix is well conditioned $W[\hatD]\sim 1$.
We therefore implement a full reorthogonalization step in our algorithm
to keep the orthogonality among the Lanczos vectors $q_{i}$.

Our algorithm requires $k$ working vectors to store the Lanczos vectors. 
However, we already employ $\Npoly$ working vectors to 
calculate the quark force in the HMC step, which we can reuse 
for the Lanczos vectors. 
A possible problem that the dimension of Krylov subspace, $k$, exceeds 
the number of working vectors, $\Npoly$, do not arise in practice 
for large $\Npoly$ because 
large $\Npoly$ means that $W$ is very close to unity so that the Lanczos 
algorithm stops at earlier steps.

We can consider various types of stopping condition.  For example,
$\mathit{err}_1$ is based on the residual in CG algorithm for 
the calculation of $A^{-1}b$~\cite{MATRIXCOMPUTATION},
and $\mathit{err}_2$ on a  
comparison of the successive approximation $x_{i}$ as described in 
Algorithm~\ref{alg:alg1}. 
In order to avoid explicit residual calculation, 
these stopping criteria should ensure that the residual
decreases to sufficient level during the iteration.
For the CG-based stopping condition, which was originally introduced
by Bori\c{c}i~\cite{Borici_matrix_function},
the analytical relation between the CG-based stopping condition and
the truncation error of the Lanczos iteration is discussed by 
van~den~Eshof \textit{et al.}~\cite{Eshof_Frommer_Lippert_Schilling_Vorst},
where it is proved that the CG-based stopping condition for the inverse 
square root in the overlap operator is a safe stopping condition.

We cannot directly apply their proof to our case because 
the exponent of the correction matrix is not limited to $-1/2$.
We employ the CG-based stopping condition $\mathit{err}_1$ 
in the noisy Metropolis test, and numerically verify the validity of this choice 
by observing the convergence behavior of the residuals and $dS$.
This analysis is described in section~\ref{sec:sec7}.
We leave the mathematical proof whether the CG stopping condition is 
safe or not for our case for future studies.

\section{Cost estimate}
\label{sec:sec6}

The computational cost of our algorithm is measured by the number of 
multiplication with $\hat{M}_{oo}$ for traversing a single unit of 
trajectory. 
The number of multiplication is separately estimated for the HMC part  
and the noisy Metropolis part.
We define the algorithmic parameters as follows:
\begin{itemize}
\item the number of MD step : $N^{A}_{\mathrm{MD}}$
\item the number of iteration of the CG algorithm : $N^{A}_{\mathrm{CG}}$
\item the order of polynomial : $\Npoly^{A}$
\end{itemize}
where the superscript $A$ refers to the case A algorithm, which should be 
replaced with $B$ for the case B algorithm. 
We use the single leapfrog scheme for the MD integrator.

\paragraph*{Cost of HMC part}

The computational cost of the HMC part of the algorithm is divided into
three pieces; calculation of the MD force with the polynomial pseudo-fermion,
the generation of pseudo-fermion field according to the polynomial action, and
the calculation of the total Hamiltonian for the HMC Metropolis test.

From Eqs.~(\ref{eq:Force}),  (\ref{eq:Memory_Force}), 
and  (\ref{eq:Recurrence_Force}) in section~\ref{sec:sec4},
the cost of the force calculation in the HMC algorithm 
is estimated as
\[
N^{A}_{\mathrm{MD}}\times (2 \Npoly^{A}-1) \mbox{\ \ \ ( case A)},
\]\[
N^{B}_{\mathrm{MD}}\times (\Npoly^{B}-1)   \mbox{\ \ \ ( case B)}.
\]
From Eqs.~(\ref{eq:PseudoFieldA}) and (\ref{eq:PseudoFieldB}),
the computational cost to generate the pseudo-fermion field is estimated as
\[
  ( (8/N_f)\times \Npoly^{A} + 1)\times N_{\mathrm{CG}}^{A}
 +(8/N_f-1)\times \Npoly^{A} + 1 
    \mbox{\ \ \ \ (case A)},
\]
\begin{eqnarray}
  ( (4/N_f)\times \Npoly^{B} + 1)\times N_{\mathrm{CG}}^{B}
 +(4/N_f-1)\times \Npoly^{B} + \Npoly^{B}/2 + 1 && \nonumber\\
  &&\mbox{\hspace{-3em}(case B)}.\nonumber
\end{eqnarray}
The computational cost of the Hamiltonian comes from the calculation 
of the pseudo-fermion action at the end of the MD step.
The number of multiplication of $\hat{M}_{oo}$ is estimated as
$\Npoly^{A}$ for the case A and $\Npoly^{B}/2$ for the case B.

Summarizing, the computational cost of the HMC part of our
algorithm is given by 
\begin{eqnarray}
N^{A}_{\mathrm{HMC\ cost}}
&=&(2\Npoly^{A}-1)\times N^{A}_{\mathrm{MD}} 
  + ( (8/N_f)\times \Npoly^{A} + 1 )\times N^{A}_{\mathrm{CG}}\nonumber\\
&&+ (8/N_f-1)\times \Npoly^{A} + 1,
\end{eqnarray}
for the case A, and
\begin{eqnarray}
N^{B}_{\mathrm{HMC\ cost}}
&=&(\Npoly^{B}-1)\times N^{B}_{\mathrm{MD}}
  + ( (4/N_f)\times \Npoly^{B} + 1 )\times N^{B}_{\mathrm{CG}}\nonumber\\
&&+ (4/N_f-1)\times \Npoly^{B} + \Npoly^{B}/2 + 1,
\end{eqnarray}
for the case B. 

We observed in section~\ref{sec:sec3}
that $\Npoly^{A}\sim \Npoly^{B}$ at a comparable value of 
the polynomial approximation residual Eq.~(\ref{eq:Residual}).
Assuming $N_{\mathrm{MD}}^{A}\sim N_{\mathrm{MD}}^{B}$
and $N_{\mathrm{CG}}^{A}\sim N_{\mathrm{CG}}^{B}$, we find 
$N^{A}_{\mathrm{HMC\ cost}}\sim 2 N^{B}_{\mathrm{HMC\ cost}}$.
We conclude that the cost of HMC part of the algorithm
is twice better for the case B than for the case A. 

\paragraph*{Cost of noisy Metropolis part}

Here we estimate the cost of the noisy Metropolis test,
$N^{A}_{\mathrm{NMP\ cost}}$ and $N^{B}_{\mathrm{NMP\ cost}}$,
for both cases.  The cost arises from Eq.~(\ref{eq:dS_corr_simple}), 
and is estimated as
\begin{equation}
N^{A}_{\mathrm{NMP\ cost}}=
  ( (8/N_f)\times \Npoly^{A} + 1) \times N^{A}_{\mathrm{CG}} \times 2,
\end{equation}
for the case A, and
\begin{equation}
N^{B}_{\mathrm{NMP\ cost}}=
  ( (4/N_f)\times \Npoly^{B} + 1) \times N^{B}_{\mathrm{CG}} \times 2,
\end{equation}
for the case B. 
The factor $2$ arises since we need to call twice the Lanczos-based algorithm
to calculate $W[\hatD]^{N_f/8}\eta_{o}$ and $W[\hatD']^{N_f/4}\zeta_{o}$.
Here we used $N_{\mathrm{CG}}^{A}$ ($N_{\mathrm{CG}}^{B}$)
as the number of Lanczos iteration. This is because
the number of Lanczos iteration is expected to be almost identical 
to that of CG iteration to generate the pseudo-fermion field,
when we employ the CG based stopping criterion.
We expect
$N^{A}_{\mathrm{CG}}\sim N^{B}_{\mathrm{CG}}$ and $\Npoly^{A}\sim\Npoly^{B}$
as before.  
Then we find
$N^{\mathrm{NMP}}_{\mathrm{cost\ A}}\sim 2 N^{\mathrm{NMP}}_{\mathrm{cost\ B}}$.

\paragraph*{Total cost}

Combining the result on the computational cost for the HMC step and that for the noisy 
Metropolis test described above,
we find that the cost for
the case A algorithm is larger than that for the case B algorithm by a factor two when
the approximated polynomials for the two algorithms have the same residual on the correction
matrix.  In our numerical test described in the next section, we 
employ the case B algorithm.

\section{Numerical tests}
\label{sec:sec7}

We test our algorithm (case B) with the Chebyshev polynomial 
on three lattices in the two-flavor case.
The simulation program is written in the optimized FORTRAN90 on SR8000 
model F1 at KEK.
Double precision arithmetic is applied to the whole numerical operations.
The lattice volume, gauge coupling, and quark masses we employed 
are shown in Table~\ref{tab:Params}.
The ``small size'' lattice is used to investigate the basic property
of the algorithm.
We show the $\Npoly$ dependence of $\langle dS\rangle$ and the 
averaged acceptance rate of the noisy Metropolis test 
$\langle P_{\mathit{corr}}\rangle$. 
The $\Npoly$ dependence of averaged
plaquette $\langle P\rangle$ is also presented on this lattice, and a comparison to results 
from the $R$-algorithm is also shown.
Using the ``middle size'' lattice, we compare the $\Npoly$ dependence 
of $\langle dS\rangle$ for different quark masses.
We employ the ``large size'' lattice parameter to see the
applicability of our algorithm to realistic simulations,
where we check the reversibility of the MD evolution and the validity
of the CG-based stopping criterion for the Lanczos algorithm in the
noisy Metropolis test. 
A comparison of $\langle P\rangle$ to the $R$-algorithm is also made.

The unit trajectory length is chosen to be $1$ in the following.
We employ the single leapfrog scheme for the  MD evolution, and
call the number of MD step as $N_{\mathrm{MD}}$. The parameter  
$a\Lambda_{\mathrm{max}}$ is roughly optimized during the thermalization 
period for each lattice parameter.

\begin{table}[b]
  \begin{center}
  \caption{Simulation parameter}
  \label{tab:Params}
    \begin{tabular}{|c|ccc|} \hline
             & volume           &$\beta$& $am$          \\\hline
  Small size & $8^3\times 4$    & 5.26  & 0.025         \\
 Middle size & $8^3\times 16$   & 5.70  & 0.01 and 0.02 \\
  Large size & $16^4$           & 5.70  & 0.02          \\\hline
    \end{tabular}
  \end{center}
\end{table}

\subsection{Results on the small size lattice}

\begin{figure}[t]
  \begin{center}
    \includegraphics[scale=\figscale]{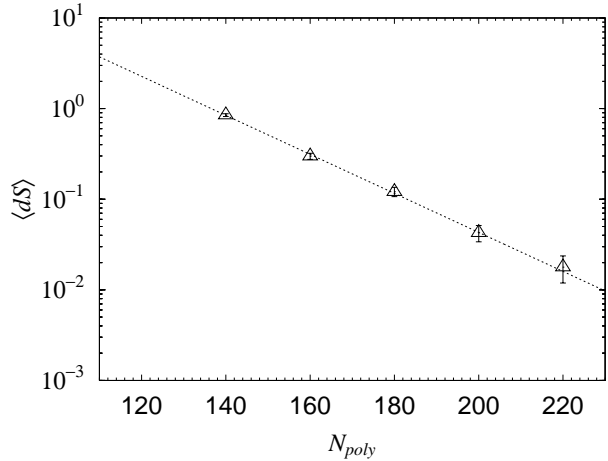}
    \caption{$\Npoly$ dependence of $\langle dS\rangle$
    on the small size lattice.}
    \label{fig:dSvsNpolyonSmall}
  \end{center}
\end{figure}

On the small size lattice, $a\Lambda_{\mathrm{max}}$ is chosen to 
be $2.37$.
Figure~\ref{fig:dSvsNpolyonSmall} shows the $\Npoly$ dependence
of $\langle dS\rangle$ for this lattice, where
the number of MD step is $N_{\mathrm{MD}}=25$. 
The dotted line shows a two-parameter fit to 
$a \exp(-b\Npoly)$. We observe a clear exponential decay as expected.

\begin{figure}[t]
  \begin{center}
    \includegraphics[scale=\figscale]{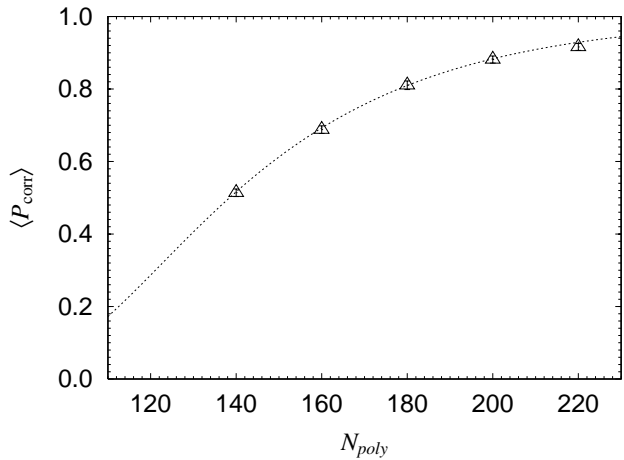}
    \caption{$\Npoly$ dependence of noisy Metropolis test acceptance rate
    on the small size lattice.}
    \label{fig:GAccvsNpolyonSmall}
  \end{center}
\end{figure}

Figure~\ref{fig:GAccvsNpolyonSmall} shows the $\Npoly$ dependence of 
the averaged noisy Metropolis acceptance rate $\langle P_{\mathit{corr}}\rangle$.
We observe consistent results to
the theoretical ansatz $\mathrm{erfc}(\left(a \exp(-b\Npoly)\right)^{1/2}/2)$,
where the dotted curve shows the ansatz with $a$ and $b$ obtained in 
Figure~\ref{fig:dSvsNpolyonSmall}.

\begin{figure}[t]
  \begin{center}
  \includegraphics[scale=\figscale]{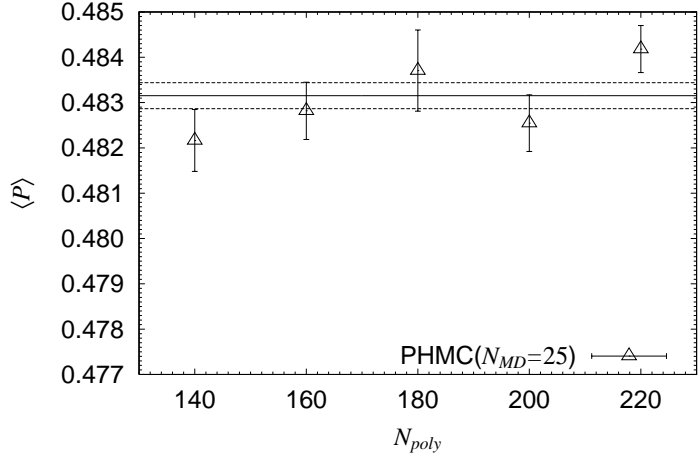}
  \caption{$\Npoly$ dependence of the averaged plaquette on small 
     size lattice.}
  \label{fig:PlaqvsNpolyonSmall}    
  \end{center}
\end{figure}

\begin{figure}[t]
  \begin{center}
  \includegraphics[scale=\figscale]{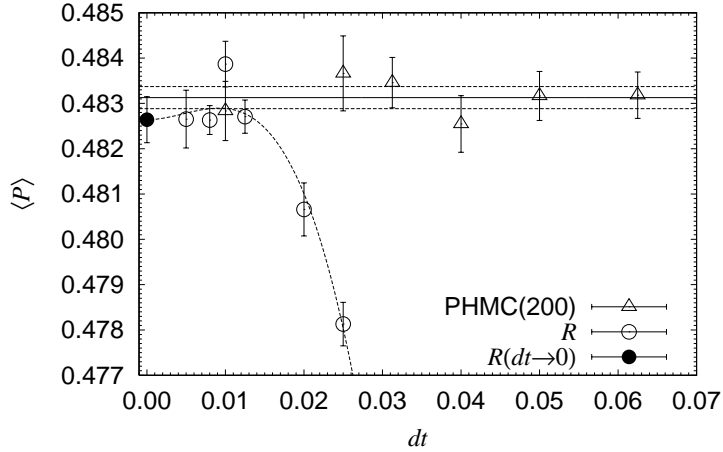}
  \caption{MD step size $dt$ dependence of the averaged plaquette
           on the small size lattice.
           The dotted curve shows the fit with $f(dt)=a dt^2 + b dt^3+c$
           for the results with $R$-algorithm.}
  \label{fig:CompHR}    
  \end{center}
\end{figure}

We show the $\Npoly$ dependence of the averaged plaquette $\langle P\rangle$
in Figure~\ref{fig:PlaqvsNpolyonSmall}. 
The horizontal line shows the constant fit to the results. The fit error
is presented by the dashed lines.
The results does not depend on $\Npoly$ as it should be.
In Figure~\ref{fig:CompHR},
we plot the MD step size, $dt=1/N_{\mathrm{MD}}$, dependence of 
the averaged plaquette together with the results with the $R$-algorithm.
We employ $\Npoly=200$ for the PHMC algorithm in the figure.
Since the $R$-algorithm has $O(dt^2)$ errors, we fit the results with 
the $R$-algorithm with $f(dt)=a dt^2 + b dt^3+c$ as shown by the dotted curve.
The horizontal lines show the constant fit to the PHMC results again.
The result with the PHMC algorithm does not depend on $dt$ and 
is consistent with that of the zero step size limit of the $R$-algorithm.
With these observations we conclude that the PHMC algorithm works 
perfectly on the small size lattice.


\subsection{Results on the middle size lattice}

\begin{figure}[t]
  \begin{center}
  \includegraphics[scale=\figscale]{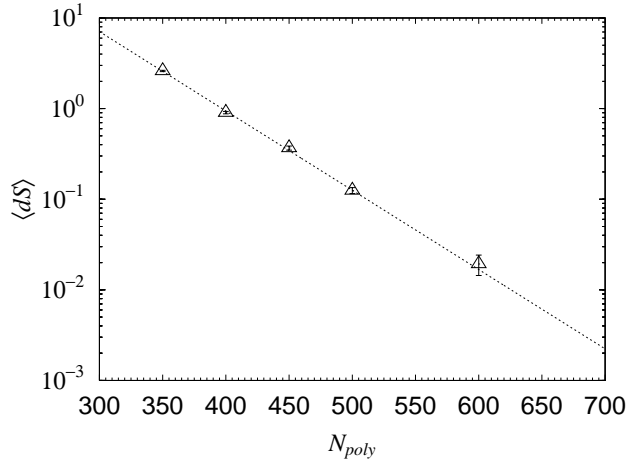}
  \caption{Same as Figure~\ref{fig:dSvsNpolyonSmall}, but 
           for the middle size lattice with $am=0.01$.}
  \label{fig:dSvsNpolyonSmallLight}    
  \end{center}
\end{figure}

\begin{figure}[t]
  \begin{center}
  \includegraphics[scale=\figscale]{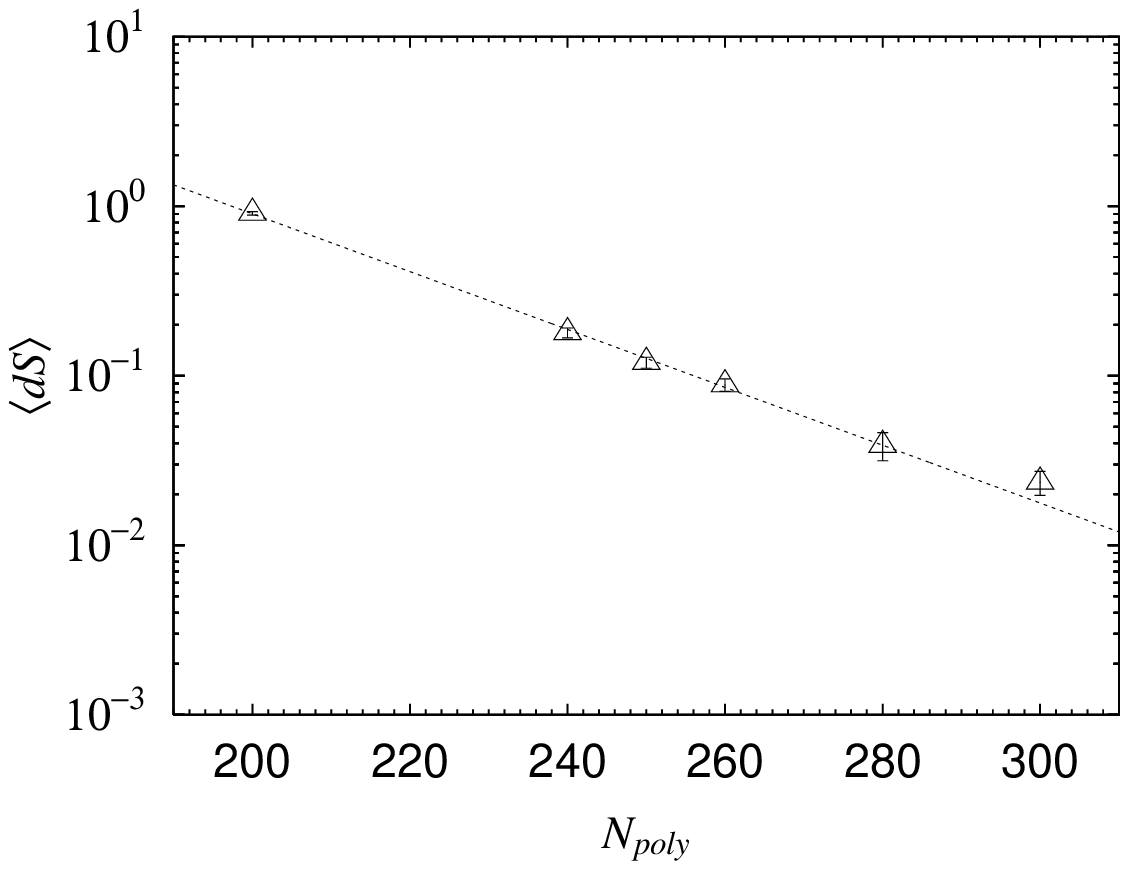}
  \caption{Same as Figure~\ref{fig:dSvsNpolyonSmall}, but 
           for the middle size lattice with $am=0.02$.}
  \label{fig:dSvsNpolyonSmallHeavy}    
  \end{center}
\end{figure}

We plot the $\Npoly$ dependence of $\langle dS\rangle$ for 
$am=0.01$ and $0.02$ in 
Figs.~\ref{fig:dSvsNpolyonSmallLight}
and \ref{fig:dSvsNpolyonSmallHeavy}, respectively.
Here $N_{\mathrm{MD}}=50$ and $a\Lambda_{\mathrm{max}}=2.28$ are 
used for both masses.
A clear exponential decay is observed in both figures. 
These behaviors are similar to those seen for the small size lattice. 

For the $am=0.01$ case, which corresponds to the ratio
of pseudo-scalar and vector meson masses 
$m_{\mathrm{PS}}/m_{\mathrm{V}}\sim 0.61$~\cite{JLQCD_KS},
we need $\Npoly\sim$ 500-600 for reasonable acceptance rate for 
the noisy Metropolis test. 
Moreover, 
assuming $a\Lambda_{\mathrm{max}}/am \ll 1$ and independence 
of $a\Lambda_{\mathrm{max}}$ from $am$,
we expect that $\Npoly$ behaves as $a\Lambda_{\mathrm{max}}/am$ 
in order to keep the residual at a constant level
(see below Eq.~(\ref{eq:ErrorBound})).
This leads us to suspect that for simulations with much lighter 
quark masses a polynomial of order $O(1000)$ is required.

Our PHMC algorithm has several problems
with such a large order polynomial. One is the memory cost in the 
calculation of the MD force from the polynomial pseudo-fermion.
Fortunately this would not be a serious hindrance in nowadays 
high performance computing 
since memory cost is relatively low compared to the Wilson fermions
(the KS fermions do not have spin indices).
Another problem is the extraction of the polynomial coefficients 
of $Q_{\Npoly}$ from the original polynomial $P_{\Npoly}$ 
as described in section~\ref{sec:sec2}.


\begin{table}[b]
  \begin{center}
    \caption{Numerical results with PHMC ($am=0.02$, $a\Lambda_{\mathrm{max}}=2.28$) 
    on the large size lattice.
    $\langle P\rangle$ : averaged plaquette.
    \#Mult$/$traj : averaged
      number of multiplication of $M_{eo}$ and $M_{oe}$ to achieve unit trajectory.
    }
    \label{tab:ResultsonLarge}
  \begin{tabular}{|c|ccc|c|}\hline
   $N_{\mathrm{poly}}$ &        300 &        400 &        500 & R algorithm~\cite{JLQCD_KS} \\
$[dt,N_{\mathrm{MD}}]$ &$[0.02,50]$ &$[0.02,50]$ &$[0.02,50]$ &$[0.02,50]$  \\\hline
   Trajectories        &  1700      &   1050     &    800     &  300        \\
   \#Mult$/$traj       &  61291(183)&  73176(296)&  87955(350)&  -          \\\hline
    $\langle P\rangle$ &0.577099(46)&0.577130(46)&0.577023(43)& 0.577261(49)\\
   \begin{tabular}{c}
            HMC \\
            Acceptance 
   \end{tabular}       & 0.8059(103)& 0.7962(168)& 0.7775(194)&  -          \\
   $\langle dH\rangle$ & 0.1112(126)& 0.1359(147)& 0.1497(187)&  -          \\
   \begin{tabular}{c}
            Correction \\
            Acceptance 
   \end{tabular}       & 0.7837(128)& 0.9627(70) & 0.9871(45) &  -          \\
   $\langle dS\rangle$ & 0.1331(164)&-0.0002(29) & 0.0000(6)  &  -          \\
   \begin{tabular}{c}
            Total \\
            Acceptance
   \end{tabular}       & 0.6329(122)& 0.7657(168)& 0.7675(191)&  -          \\\hline
    \end{tabular}
  \end{center}
\end{table}

\begin{figure}[t]
  \begin{center}
  \includegraphics[scale=\figscale]{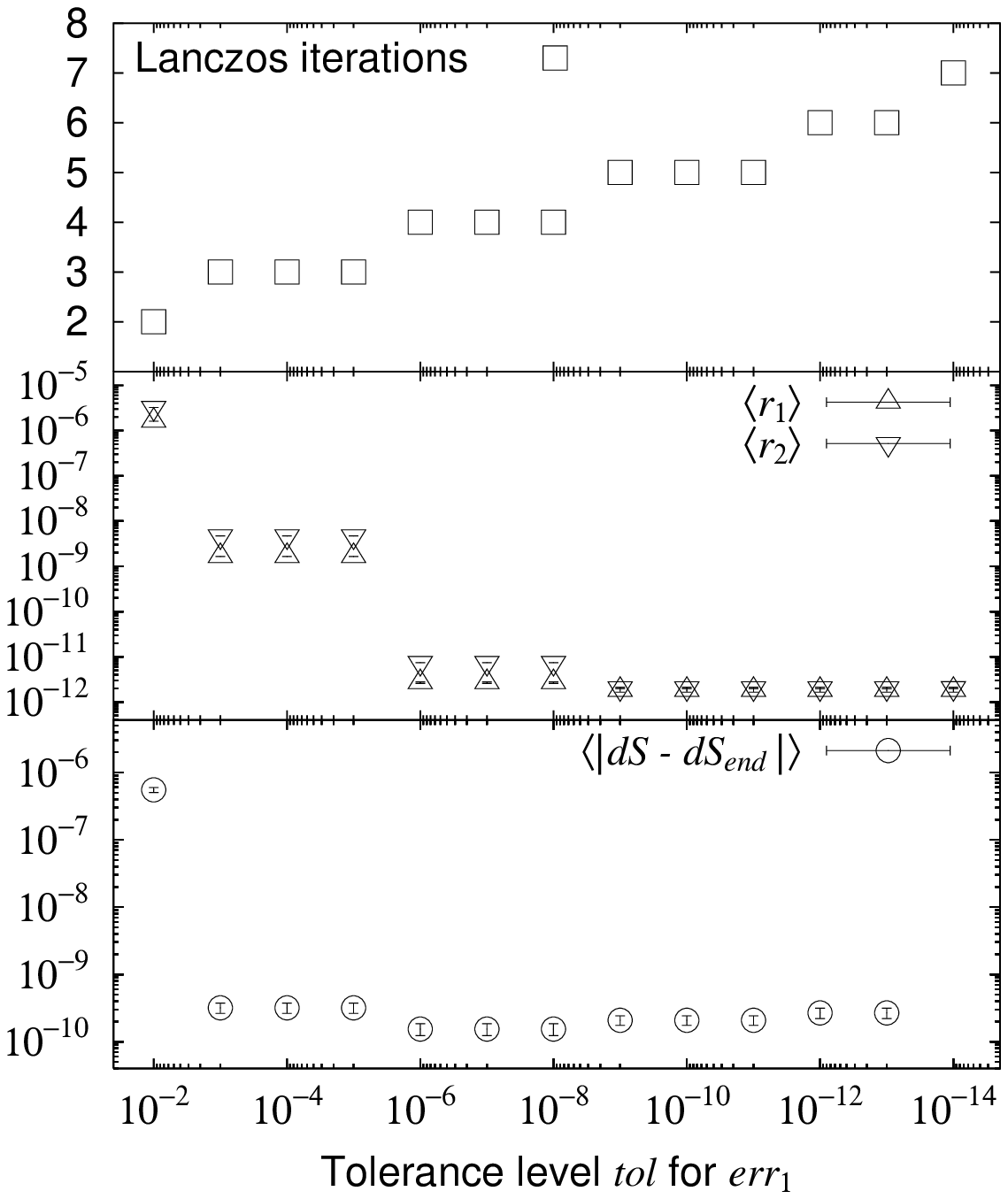}
  \caption{The convergence behavior of the Lanczos-based algorithm
           for $dS$ calculation as a function of the tolerance level.
           Upper figure shows the number of iteration in the Lanczos algorithm,
           middle one shows the residuals defined in Eqs.~(\ref{eq:Res1}) and 
           (\ref{eq:Res2}) for $(W[\hatD])^{N_{f}/8}$ and ${(W[\hatD'])}^{-N_{f}/4}$,
           bottom one for the $|dS-dS_{\mathit{end}}|$ where $dS_{\mathit{end}}$ is
           the value of $dS$ itself at $\mathit{tol}=10^{-14}$.}
  \vspace*{1em}
  \label{fig:TolvsdS_L}    
  \end{center}
\end{figure}

\subsection{Results on the large size lattice}

We show the results on the large size lattice in Table~\ref{tab:ResultsonLarge}.
We quote the averaged plaquette value with the $R$-algorithm from Ref.~\cite{JLQCD_KS}.
We observe a roughly $2\sigma$ deviation, 
which may be ascribed to a finite step size error inherent in the $R$-algorithm. 
The number of multiplication of hopping matrices $M_{eo}$ and $M_{oe}$ measured in 
the program is roughly proportional to $\Npoly$, which is expected
from the discussion in section~\ref{sec:sec6}. 
We will discuss the finite step size error of the $R$-algorithm and
compare the computational cost of the PHMC algorithm to that of the $R$-algorithm 
after describing the numerical property of the reversibility violation, the Lanczos 
algorithm, and $dS$ for the PHMC algorithm.

As described in section~\ref{sec:sec4}, we investigate the reversibility 
violation on the large size lattice using the following observables:
\begin{eqnarray}
  \Delta H / H & \equiv & |H(t_{r}-t_{r})-H(0)|/H(0),\\
  \Delta U &\equiv & \sqrt{\sum_{n,\mu,a,b}\left|(U_{\mu})_{a,b}(n)(t_{r}-t_{r})
-(U_{\mu})_{a,b}(n)(0)\right|},\\
  \Delta P &\equiv & \sqrt{\sum_{n,\mu,a,b}\left|(P_{\mu})_{a,b}(n)(t_{r}-t_{r})
-(P_{\mu})_{a,b}(n)(0)\right|},
\end{eqnarray}
where $X(0)$ means the observable $X$ calculated at the initial configuration
at $t=0$ in the MD evolution and 
$X(t_{r}-t_{r})$ the observable calculated at  
the reversed configuration which is obtained from the initial 
configuration at $t=0$ by integrating the MD equation to $t=t_{r}$ 
and then integrating back to $t=0$. The trajectory length is $t_{r}=1$.
We measured these quantities using 20 configurations which are separated by
5 trajectories. We observe 
\begin{eqnarray}
  \langle \Delta H / H \rangle &=& 
     0.26(6) \times 10^{-15},\nonumber \\
  \langle \Delta P\rangle/\sqrt{9\times 4\times N_{\mathrm{vol}}}&=&
     0.4162(7) \times 10^{-14},\nonumber \\
  \langle \Delta U\rangle/\sqrt{9\times 4\times N_{\mathrm{vol}}}&=&
     0.1484(2) \times 10^{-14},
\end{eqnarray}
with the $\Npoly=300$ PHMC algorithm. The errors are estimated
with the binned jackknife method.
These values are at a completely satisfactory level with the double precision 
arithmetic. 
Consequently it is concluded that
the method we employed for the force calculation of the polynomial 
pseudo-fermion is stable in the MD evolution and does not cause 
violation of reversibility.

The validity of the CG based stopping criterion for the Lanczos method
to calculate $dS$ in the noisy Metropolis test is also investigated on 
the same 20 configurations. We measured $dS$, and two residuals defined by
\begin{eqnarray}
  \label{eq:Res1}
  r_{1}&=& \left| \left((W[\hatD])^{N_f/8}\right)^{8/N_f}\eta_{o}-W[\hatD]\eta_{o}\right|/
           |W[\hatD]\eta_{o}|,\\
  \label{eq:Res2}
  r_{2}&=& \left| W[\hatD']\left((W[\hatD'])^{-N_f/4}\right)^{4/N_f}\zeta_{o}-\zeta_{o}\right|/
           |\zeta_{o}|,
\end{eqnarray}
where $\zeta_{o}$ is defined in Eq.~(\ref{eq:dS_corr_simple}),
and the number of iteration of the Lanczos iteration by varying the tolerance 
level $\mathit{tol}$ for $err_{1}$.

Figure~\ref{fig:TolvsdS_L} shows the convergence behavior of the above quantities.
The number of iteration increases step by step with the decreasing stopping 
condition (top of the figure). 
The residuals Eqs.~(\ref{eq:Res1}) and (\ref{eq:Res2}) stagnate around 
a $O(10^{-12})$ level (middle of the figure). 
As the residuals are defined as relative error, one may suspect that 
the number of $O(10^{-12})$ is not sufficient with the double precision arithmetic. 
We think this is due to the accumulation of round-off errors in  applying 
the Lanczos iteration several times to calculate the residuals.
Namely, for the explicit residual calculation, we need four times application of 
the Lanczos iteration for $r_{1}$ and twice for $r_{2}$ in the $N_f=2$ case,
and the Lanczos iteration does not span the completely same Krylov subspace
in each application.

For the correctness of the algorithm, $dS$ itself is more important.
To see the stopping condition dependence of $dS$, we measured 
the convergence of $|dS-dS_{end}|$ as the metric and show the result
in the bottom of Figure~\ref{fig:TolvsdS_L}, where $dS_{end}$ is $dS$ at 
the most stringent stopping condition $\mathit{tol}=10^{-14}$. 
Since the change is too rapid against the change of the number of iteration, 
we could not see the exponential decay. 
The metric stagnates around $O(10^{-10})$. The reason is understood as follows.
$dS$ is defined as the difference of $\zeta^{\dag}_{o}W[\hatD']^{-N_f/4}\zeta_{o}$ 
and $|\eta_{o}|^2$ in Eq.~(\ref{eq:dS_corr_simple}).
Numerically it is observed that
$\zeta^{\dag}_{o}W[\hatD']^{-N_f/4}\zeta_{o}$ and $|\eta_{o}|^2$.
have almost the same values of about $3\times 16^{4}/2\sim O(10^{5})$, 
and the resulting $dS$ is of $O(10^{-1})$.
Within double precision arithmetic, the subtraction of $O(10^{5})$ 
from $O(10^{5})$ yielding $O(10^{-1})$ for $dS$ 
means that $dS$ only has 9-10 significant figures.
The stagnation around $O(10^{-10})$ would occur in such a case.
We stress that the 9-10 significant figures for $dS$ is sufficient
in the realistic simulations with $O(10^4)$ trajectories.
No visible effect from the choice of the CG-based stopping criterion is observed.

The plaquette value of the $R$-algorithm from Ref.~\cite{JLQCD_KS} 
is larger than that of the PHMC algorithm about $2\sigma$ as shown in 
Table~\ref{tab:ResultsonLarge}. 
This may contradict the previous observation on the small size lattice 
where the plaquette value of the $R$-algorithm approaches the value of 
the PHMC algorithm from smaller value. 
To make clear the $dt$ dependence of the plaquette with the $R$-algorithm, 
we imported the program of the $R$-algorithm into SR8000-F1 model and 
produced several trajectories with the $R$-algorithm on the large size 
lattice.


\begin{table}[b]
  \begin{center}
    \caption{The results with the $R$-algorithm on the large size lattice 
             ($16^{4}, \beta=5.7, am=0.02$). 
             $\mathit{res}$ is used for the stopping criterion of 
             the CG algorithm in the $R$-algorithm.}
    \label{tab:HR_LL}
    \begin{tabular}{|c|cccc|}\hline
    $\mathit{res}$
                       & $10^{-4}$   & $10^{-4}$   & $10^{-4}$   & $10^{-4}$  \\
$[dt,N_{\mathrm{MD}}]$ & [0.005,200] & [0.01,100]  & [0.02,50]   & [0.025,40] \\\hline
           \# of Traj. &  800        & 1200        & 1200        & 1000       \\
  \#Mult/traj          & 113210(450) &  56700(355) & 28627(113)  & 23102(76)  \\\hline
    $\langle P\rangle$ &0.577102(63) &0.577133(67) &0.577194(71) &0.577294(67)\\\hline\hline
    $\mathit{res}$
                       & $10^{-4}$   & $10^{-4}$   & $10^{-4}$   & $10^{-8}$  \\
$[dt,N_{\mathrm{MD}}]$ & [0.04,25]   & [0.05,20]   & [0.0625,16] & [0.02,50]  \\\hline
           \# of Traj. & 1500        & 2000        & 2500        & 1000       \\
  \#Mult/traj          & 14432(130)  & 11808(42)   & 9478(91)    & 77880(807) \\\hline
    $\langle P\rangle$ &0.577389(71) &0.577076(100)&0.576423(232)&0.577411(67)\\\hline\hline
    $\mathit{res}$
                       & $10^{-12}$  & $10^{-15}$  & - & - \\
$[dt,N_{\mathrm{MD}}]$ & [0.02,50]   & [0.02,50]   & - & - \\\hline
           \# of Traj. & 1000        &  800        & - & - \\
  \#Mult/traj          & 127772(1162)& 166573(424) & - & - \\\hline
    $\langle P\rangle$ &0.577335(63) &0.577242(57) & - & - \\\hline
    \end{tabular}
  \end{center}
\end{table}

\begin{figure}[t]
  \begin{center}
  \includegraphics[scale=\figscale]{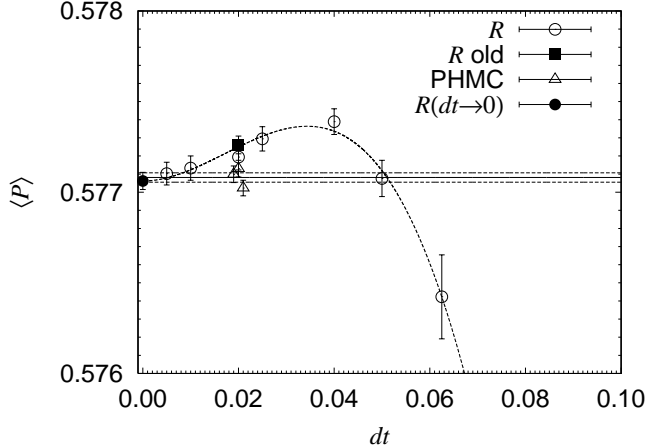}
  \caption{MD step size $dt$ dependence of the averaged plaquette on the large size
           lattice. The dotted curve shows the fit with $f(dt)=a dt^2 + b dt^3 + c$
           to open circles with the $R$-algorithm. 
           Filled square is the result from Ref.~\cite{JLQCD_KS}.}
  \label{fig:Plaqvsdt_LL}
  \end{center}
\end{figure}

Table~\ref{tab:HR_LL} shows the results with the $R$-algorithm on 
the large size lattice. 
The definition of the norm $\mathit{res}$ for the stopping criterion
of the CG algorithm required in the $R$-algorithm is the same as that 
of Ref.~\cite{JLQCD_KS} (where the symbol $r$ is used instead of $\mathit{res}$).
We do not observe clear stopping criterion dependence on $\langle P\rangle$.
Figure~\ref{fig:Plaqvsdt_LL} shows the $dt$ dependence of $\langle P\rangle$ on 
the large size lattice. 
Open circles are the results with the $R$-algorithm produced for 
the comparison. Filled square is the previous result with the $R$-algorithm~\cite{JLQCD_KS}.
Open triangles are the results with the PHMC algorithm scatted around 
$dt=0.02$ for clarity of presentation. 
We observe that the $dt$ behavior of the $R$-algorithm 
in small $dt$ region is slightly different 
from that of the small size lattice (Fig.~\ref{fig:CompHR}),
while the behavior at large $dt$ region is similar to each other.
We also observe that the value of $dt$ where $\langle P\rangle$ 
largely deviate from the value at the limit $dt\rightarrow 0$ is different
(it is $dt \sim 0.01$ in Fig.~\ref{fig:CompHR} and $dt\sim 0.04-0.05$ 
in Fig.~\ref{fig:Plaqvsdt_LL}).

Although we cannot make clear statement on the discrepancy of $dt$ behavior between 
two lattice sizes, we can say that the $dt$ dependence is affected by the physical
situation and parameters.
The reason is as follows.
The error analysis of the $R$-algorithm by $dt$ perturbation fails at 
large $dt$. 
More precisely it is said that the point where the perturbative analysis 
fails is governed by $dt/m$ with $m$ the lightest fermion mode in 
the $R$-algorithm~\cite{R_RHMC}.
We do not tune the input parameters for these two simulation sets.
It is natural that the physical lightest fermion mode is different between
the small and large size lattices.
Thus we consider that this discrepancy comes from the different physical 
parameter and situation between the small and large size lattices and
$\langle P\rangle$ with the $R$-algorithm at $dt=0.02$ is accidentally
larger than that with the PHMC algorithm in the large size lattice.
We stress that the result from the PHMC algorithm
is consistent with the limit $dt\rightarrow 0$ of the results with 
the $R$-algorithm.

The computational cost of the PHMC algorithm of $\Npoly=400$ is comparable 
to that of the $R$-algorithm with $\mathit{res}=10^{-8}$ at $dt=0.02$
in terms of \#Mult/traj on the large size lattice. 
We need to take the limit $dt\rightarrow 0$ and use sufficiently small 
$\mathit{res}$ in the $R$-algorithm theoretically, which requires significantly 
large amount of computational cost for the $R$-algorithm.
The actual computational time for the PHMC algorithm with $\Npoly=400$ on the
large size lattice was measured as $136$ sec for unit trajectory with 
SR8000-F1 4-nodes 
(peak speed : 12$\times$4 GFlops, sustained speed : 3.5$\times$ 4 GFlops) and
it is $120$ sec for the $R$-algorithm with $\mathit{res}=10^{-8}$ 
(sustained speed : 3.4$\times$ 4 GFlops) at $dt=0.02$.
The same computational speed is achieved because both programs make
use of the common subroutine for the hopping matrix multiplication.
Thus we conclude that the PHMC algorithm works on the lattice size of 
$16^{4}$ with a quark mass $am=0.02$ corresponding to 
$m_{\mathrm{PS}}/m_{\mathrm{V}}\sim 0.69$~\cite{JLQCD_KS} within 
reasonable computational time and is applicable to realistic simulations.
The advantage of exact algorithm is very clear in this situation.

\section{Conclusion}
\label{sec:sec8}

In this paper, we have presented an exact algorithm for dynamical 
lattice QCD simulation with the KS fermions where single flavor quark 
is defined as the $1/4$ power of the KS fermion matrix.
The algorithm is an extension of the polynomial Hybrid Monte Carlo (PHMC)
algorithm in which the Hermitian polynomial approximation is applied to
the fractional power of the KS fermion matrix.
The Kennedy-Kuti noisy Metropolis test is incorporated to make the
algorithm exact.

We introduced two types of polynomial approximations and corresponding
PHMC algorithms. One algorithm approximates $\hatD^{-N_f/8}$ with a single polynomial
$P_{\Npoly}[\hatD]$ (case A), 
and the other $\hatD^{-N_f/4}$ with a squared polynomial 
$|Q_{\Npoly}[\hatD]|^2$ (case B).
For the noisy Metropolis test, we made use of a Lanczos-based 
Krylov subspace method to calculate the fractional power of 
the correction matrix.
The efficiency of the two algorithms was estimated, and it was found that 
the latter (case B) had better performance than that of the former 
by a factor two.

We tested our algorithm (case B) using the Hermitian Chebyshev 
polynomial for the case of two-flavor QCD on three lattice sizes.
Results on a small lattice of $8^3\times 4$ demonstrated that the 
algorithm works correctly, {\it e.g.}, the averaged 
plaquette value agrees with that of the $R$-algorithm 
after extrapolation to the zero step size limit.
We have also shown that our algorithm works on a moderately large lattice of 
size $16^4$,  albeit for a rather heavy quark mass of 
$m_{\mathrm{VS}}/m_{\mathrm{V}}\sim 0.69$, within reasonable simulation costs
compared to that of the $R$-algorithm.

There are several points that require improvements with our work. 
One of the points concerns the fact that the calculation of the polynomial 
coefficients in case B for splitting the original polynomial becomes 
progressively difficult toward lighter quark masses. 
While this is not a limitation of the PHMC algorithm itself,
solutions should be found to solve this problem for future realistic 
simulations since the case A algorithm, which has no such problem, 
is expected to be twice slower than than the case B algorithm. 
Another point is that further improvement of the algorithm may be achieved 
by optimizing the polynomial approximation,   
and by combining the preconditioning technique and the polynomial approximation.

Anticipating progress on these fronts, we conclude that our algorithm provides 
an attractive method for dynamical KS fermion simulations for $2+1$-flavor 
QCD without systematic errors originating from the simulation algorithm. 

\section*{Acknowledgments}
This work is supported by the Supercomputer Project No.~79 (FY2002)
of High Energy Accelerator Research Organization (KEK), and also
in part by the Grant-in-Aid of the Ministry of Education 
(Nos.  11640294, 12304011, 12640253, 12740133, 13135204, 13640259, 
       13640260, 14046202, 14740173).
N.Y. is supported by the JSPS Research Fellowship.


\end{document}